\documentclass[sigconf]{acmart}
\AtBeginDocument{%
  }

\copyrightyear{2025}
\acmYear{2025}
\setcopyright{cc}
\setcctype{by}
\acmConference[CHI '25]{CHI Conference on Human Factors in Computing Systems}{April 26-May 1, 2025}{Yokohama, Japan}
\acmBooktitle{CHI Conference on Human Factors in Computing Systems (CHI '25), April 26-May 1, 2025, Yokohama, Japan}\acmDOI{10.1145/3706598.3713818}
\acmISBN{979-8-4007-1394-1/25/04}

\definecolor{revc}{RGB}{0, 0, 0} %clear
\newcommand{\rev}[1]{{\color{revc}#1}}

\definecolor{aliceblue}{rgb}{0.91, 0.94, 0.97}

\newcommand{\system}{\textsc{Amuse}}
\newcommand{\aria}{\textsc{Aria}}
\newcommand{\control}{\texttt{Assist}}
\newcommand{\baseline}{\texttt{Baseline}}

\definecolor{pBlue}{HTML}{B8E3E9}
\definecolor{vPink}{HTML}{F8C8DC}
\definecolor{eLime}{HTML}{CCEFB9}
\definecolor{pineGreen}{HTML}{01796F}

\newcommand{\dcoherent}{coherent}

\usepackage{xcolor}
\usepackage{xspace}
\usepackage{subcaption}
\usepackage{colortbl}
\usepackage{multirow}
\usepackage{amsmath}    
\usepackage{algorithm}
\usepackage{algpseudocode}
\usepackage{graphicx}
\usepackage{nicefrac}
\usepackage{framed}
\usepackage{array}
\usepackage{makecell}

\makeatletter
\@ifpackageloaded{arydshln}{
   \newcommand{\dashedline}[1]{
    \cdashline{#1}[.4pt/1pt]\noalign{\vskip 1pt}
  }
}{
  \newcommand{\dashedline}[1]{
    \cmidrule[.0025mm]{#1}
  }
}
\makeatother

\begin{document}

\title[\system{}: Human-AI Collaborative Songwriting with Multimodal Inspirations]{\system{}: Human-AI Collaborative Songwriting \\with Multimodal Inspirations} 

\author{Yewon Kim}
\authornote{Work mainly done at Carnegie Mellon University as a visiting researcher.}
\affiliation{%
  \institution{KAIST}
  \city{}
  \country{Republic of Korea}}
\email{yewon.e.kim@kaist.ac.kr}

\author{Sung-Ju Lee}
\affiliation{%
  \institution{KAIST}
  \city{}
  \country{Republic of Korea}
}
\email{profsj@kaist.ac.kr}

\author{Chris Donahue}
\affiliation{%
  \institution{Carnegie Mellon University}
  \city{}
  \country{United States}
}
\email{chrisdonahue@cmu.edu}


\begin{abstract}
Songwriting is often driven by multimodal inspirations, such as imagery, narratives, or existing music, yet songwriters remain unsupported by current music AI systems in incorporating these multimodal inputs into their creative processes.
We introduce \system{}, a songwriting assistant that transforms multimodal~(image, text, or audio) inputs into chord progressions that can be seamlessly incorporated into songwriters' creative process. 
A key feature of \system{} is its novel method for generating coherent chords that are relevant to music keywords in the absence of datasets with paired examples of multimodal inputs and chords. 
Specifically, we propose a method that leverages multimodal LLMs to convert multimodal inputs into noisy chord suggestions and uses a unimodal chord model to \rev{filter the suggestions.}
A user study with songwriters shows that \system{} effectively supports transforming multimodal ideas into coherent musical suggestions, enhancing users' agency and creativity throughout the songwriting process.
\end{abstract}






\maketitle

\begin{figure*}[t!]
  \includegraphics[width=\textwidth]{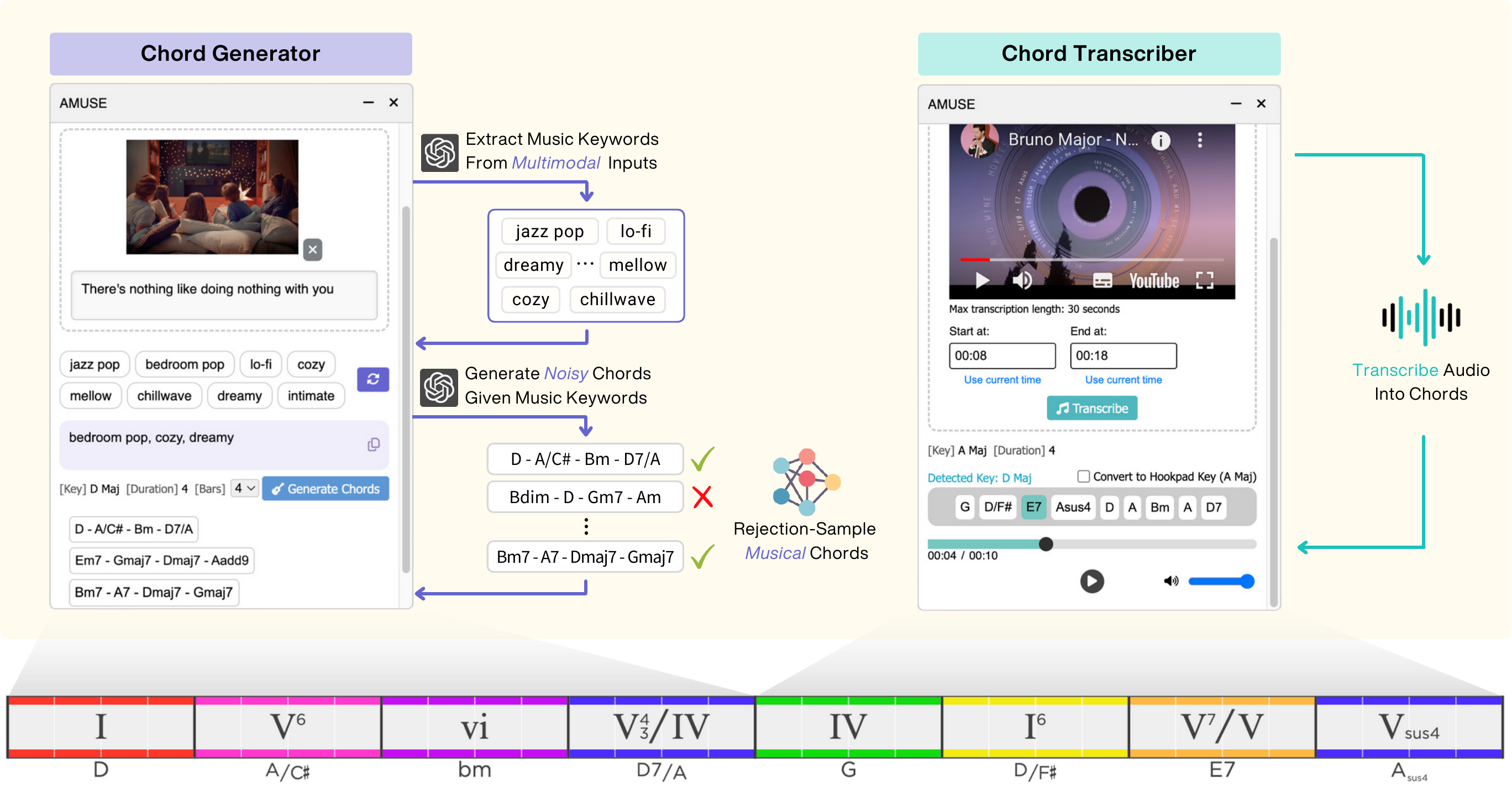}
  % \vspace{-.7em}
  \caption{\system{} transforms multimodal (image, text, or audio) inspirations into reusable musical elements (chord progressions) that songwriters can seamlessly incorporate into their creative process. \system{} consists of two functionalities: Chord Generator (Left) and Chord Transcriber (Right). In the Chord Generator, user can generate music keywords from image/text inputs and generate musically coherent chord progressions based on the music keywords. These suggestions are generated by rejection-sampling the LLM-generated chord progressions using a unimodal chord model. The Chord Transcriber allows users to transcribe chords from a specified range of audio.}
  % \vspace{-.5em}
  \Description{The overview of Amuse system, which has two main features: the Chord Generator and the Chord Transcriber. The Chord Generator extracts music keywords from inputs (image, text, or audio) to generate noisy chord suggestions, which are then refined through rejection sampling. The Chord Transcriber allows users to select a time range from audio and transcribe it into chords, which can be displayed and converted into a chosen key.}
  \label{fig:main}
\end{figure*}
\section{Introduction}

Across creative pursuits, artists draw inspiration from a wide variety of concepts and experiences, 
relying on intuition to find connections between sometimes seemingly unrelated modalities. 
Within music, for example, 
Modest Mussorgsky composed \emph{Pictures at an Exhibition} with direct inspiration from a series of paintings, 
and Bj\"ork's album \emph{Biophilia} was inspired by ruminations on the relationships between music, nature, and technology. 
Increasingly, AI systems are capable of modeling such \emph{multimodal} relationships, i.e.,~to understand not only 
\emph{concrete} relationships between data modalities (e.g.,~\textit{what objects are in this painting?}) 
but also more \emph{abstract} ones (e.g.,~\textit{what music might fit with this painting?}). 
Consequently, these systems are \rev{being applied across a growing range of creative pursuits, }
such as illustration~\citep{opal}, creative writing~\citep{talebrush}, and  video editing~\citep{wang2024lave}.

Within the context of AI support for musical creativity, recent research has centered around two key pathways. 
First, AI systems \rev{have become capable of} generating music audio from text descriptions~\citep{jukebox, riffusion, musiclm, musicgen, mousai, noise2music}, 
leveraging the particularly intuitive multimodal relationship between music and text to offer users a compelling new form of musical control. 
However, because these systems output completed songs in audio rather than \emph{reusable musical elements} (e.g.,~sequences of notes or chords), they are not particularly supportive to musicians who view \emph{iteration} as central to both the creative process and to their sense of ownership over the outcome~\citep{newman2023human}. 
In contrast, symbolic music generation systems~\citep{simon2017performance,musictransformer,payne2019musenet,donahue2019lakhnes,anticipatory} are increasingly being incorporated into music editors 
to aid musicians in iterative composition workflows by generating reusable musical elements that agree with the surrounding musical context~\citep{pianoinpaintingapplication,aria}. 
While musicians find these types of systems helpful, these systems fall short in incorporating multimodal sources of inspiration, as they only take the surrounding musical context as input. 
Accordingly, musicians remain unsupported in incorporating multimodal sources of inspiration into their iterative creative processes.

In this work, we bridge this gap by building a generative musical support tool that can both incorporate multimodal inputs and facilitate creative iteration.
Specifically, we present \system{} (Figure~\ref{fig:main}), a songwriting assistant that transforms multimodal user inputs (images, text, audio) into reusable musical elements (chord progressions)\footnote{Sound examples and code are available at: \href{https://yewon-kim.com/amuse}{\textcolor{pineGreen}{\texttt{https://yewon-kim.com/amuse}}}}.
Motivated by a formative study with eight songwriters, \system{} comprises two key functionalities: Chord Generator and Chord Transcriber. In the Chord Generator, we adopt a two-step chord generation approach where we (i) extract music keywords from user multimodal inputs and (ii) generate chord progressions that align with the given music keywords. The Chord Transcriber allows users to transcribe chords from a specified range of audio sources.
\system{} is seamlessly integrated into Hookpad~\citep{hookpad}, a music editor that assists songwriters in the composition of the chords and melody that together characterize the basic essence of a piece of music. 
In combination with \aria{}~\citep{aria}---a \emph{contextual} and \textit{unimodal} AI songwriting assistant already integrated into Hookpad---\system{} can support songwriters by incorporating multimodal sources of inspiration in addition to existing musical context.

% **Formative study**
% The design of \system{} is motivated by findings from a formative study conducted with eight songwriters who are regular users of Hookpad. 
% Our formative study was structured with the goals of both learning more about their existing songwriting processes, and exploring ways that AI might support their processes. 
% Through these conversations, we confirmed that songwriters do indeed draw inspiration from multimodal sources, 
% especially during the initial ideation phase. 
% Additionally, we learned that translating these ideas into concrete, reusable musical elements represents a challenging obstacle for musicians, and one where they would be receptive to the assistance of AI systems.

A key challenge in developing \system{} is the lack of obvious sources of training data consisting of paired multimodal inputs and reusable musical elements (i.e., chords). 
Text-to-music systems are trained on large corpora of text captions and music audio; however, to the best of our knowledge, no such corpora exist that include reusable musical elements as opposed to music audio. 
To build such a system without paired training data, we leverage the general capabilities of multimodal large language models (LLMs) by instructing GPT-4o~\cite{gpt4o} to generate chord progressions that might correspond to multimodal inputs. 
However, we find that chord progressions from GPT-4o are not musically \dcoherent{} enough 
to directly serve users. 
Equipped with a \emph{unimodal} dataset consisting only of human-composed chord progressions, 
we train a language model prior over chords, which we use to filter out unrealistic chords generated by GPT-4o through a rejection sampling procedure. Through both quantitative and qualitative evaluation, we show that our method can generate diverse chord progressions that are relevant to user keywords and musically coherent.

% **User study**:
% key findings
To understand how \system{} supports the songwriting process, we conducted a within-subjects study with 10 songwriters, where participants wrote 8-bar choruses based on songwriting prompts. Participants engaged in two conditions: one where they used both \system{} and \aria{}, and another where they used \aria{} alone. Our study revealed that \system{} effectively supports the transformation of multimodal inspirations into concrete musical elements, with participants feeling more guided and aligned with their creative goals than \aria{} alone. 
Notably, although \system{} was designed for early-stage use, participants employed it across various stages of the songwriting process, demonstrating diverse usage patterns. With \system{}, participants experienced enhanced agency and creativity throughout the process. Overall, these findings suggest that \system{} enriches the songwriting process by helping songwriters seamlessly integrate diverse multimodal inspirations into their creative workflows.

% Contributions
In summary, this paper presents the following contributions:
\begin{enumerate}
    \item \system, an interactive songwriting assistant that supports users in transforming multimodal inspirations into reusable musical elements.
    \item A novel method for generating diverse, relevant, and coherent chord progressions based on multimodal user inputs without paired training data. 
    \item Findings from a user study (N=10) on how and when multimodal inspirations are used throughout the songwriting process and their impact on the creative workflow.
\end{enumerate}
\vspace{1em}

\section{Related Work}

\subsection{Human-AI Collaborative Music Creation}

Research on human-AI collaborative music creation has largely centered on the symbolic music domain, where musical elements such as melodies and chords are represented as a sequence of symbols (e.g., MIDI). 
Most systems in this symbolic space employ \textit{contextual} models~\citep{musictransformer, anticipatory, coconet, pianoinpaintingapplication, bachdoodle}, which generate continuation or infillation given user-provided musical elements.
For example, Music Transformer~\citep{musictransformer} extends user-provided MIDI sequences by generating contextually coherent MIDI continuations. 
In addition, chord recommendation systems such as ChordRipple~\citep{chordripple} and ChordSequenceFactory~\citep{chordfactory} provide suggestions for chord progressions based on the current harmonic context, helping users build coherent and creative chord sequences over time.

In practice, human-AI collaborative music creation is facilitated by musical user interfaces, such as MIDI editors, where users can input and refine musical elements while collaborating with AI~\citep{deepbach, cococo, calliope, bachdoodle, malandro2023composer, beckerdesigning, socialglue}. A notable example is Cococo~\citep{cococo}, where users can manually input or edit musical notes, steer the contextual AI model to generate infills for incomplete sections and integrate AI-generated suggestions directly within the editor. 
These interfaces enable users to iteratively generate and refine AI suggestions, giving them agency in shaping the final output by alternating between manual input and AI suggestions~\citep{newman2023human}.
Commercial tools exemplifying these human-AI collaborative creation interfaces include \aria~\citep{aria}, which operates unimodally by generating musical elements—chords or melodies—based on user-provided contextual inputs written in the music editor.
However, prior interfaces primarily focus on \textit{contextual} musical inputs without considering other modalities (e.g., images, text) that can influence the creative process. Our study expands this scope by incorporating \textit{multimodal} inputs into these \textit{contextual} music co-creation workflows.

\subsection{Text-to-Music Generative Models}
While symbolic music AI focuses on modeling the representation of \textit{unimodal} music data, one emerging area of research involves text-to-music generative models~\citep{jukebox, riffusion, musiclm, musicgen, mousai, noise2music}. These models learn \textit{multimodal} representation between textual descriptions and music audio, allowing users to steer the model generations by text prompts (e.g., ``smooth piano improvisation over a walking bass line'').
Similarly, proprietary models like Suno~\citep{suno} and Udio~\citep{udio} allow users to input text prompts and full lyrics to generate complete songs. 
While these text-conditioned generative models offer users greater control and agency over the generation process, a key limitation is that the outputs are fully-rendered audio files, which are not easily integrated into the iterative workflows of music creation, such as those in digital audio workstations (DAWs). In contrast, our work focuses on human-AI collaborative music creation, where AI-generated outputs are meant to be part of an evolving, iterative creation process rather than final products. 

\subsection{Multimodality in Creativity Support Tools}

Multimodal systems have been widely explored in HCI and creativity support tools. In the visual domain, different modalities are used to enhance the creation of both 2D~\citep{sharma2018chatpainterimprovingtextimage, elnouby2019telldrawrepeatgenerating, opal} and 3D~\citep{Chang2017SceneSeer3S, 10.1145/383259.383316, 3dalle, 10.1145/2501988.2502008} visuals.
For instance, Opal~\citep{opal} supports the illustration creation process by incorporating inputs like news articles, keywords, tones, and styles, while in Attribit~\citep{10.1145/2501988.2502008}, users can input adjectives and emotional cues to generate 3D shapes.
Multimodal systems have also been developed to assist with writing~\citep{10.1145/3613904.3642320, 10.1145/3379503.3403556, talebrush, xcreation}. For example, PandaLens~\citep{10.1145/3613904.3642320} combines audio, gaze patterns, images, and verbal comments to generate narratives, while TaleBrush~\citep{talebrush} allows for both text-based and sketch-based interactions in storytelling. Additionally, some writing systems provide multimodal feedback rather than multimodal input to support the creative process~\citep{fairytailor, elephant}. For example, Fairytailor~\citep{fairytailor} retrieves sequences of images aligned with texts to stimulate ideation in storytelling.
Despite these advancements, multimodal music creation remains a relatively unexplored and emerging area in creativity support tools. With \system{}, we focus on integrating both contextual and multimodal inputs to explore how diverse communication channels can enhance human-AI collaborative music creation.
\section{Formative Study}
\label{section:formative}

To inform the design of \system{}, we conducted formative interviews with songwriters with two goals. Firstly, to learn about songwriters' initial ideation and song development process. Secondly, to explore potential AI supports that could enhance the songwriting process.

\subsection{Participants and Procedure}
\label{section:formative:apparatus}

We recruited eight participants who are active users of Hookpad~\citep{hookpad} and \aria~\citep{aria, ariaweb}. Hookpad is a widely-used online songwriting tool developed by Hooktheory~\cite{hooktheory} that enables users to create songs (chords and melodies) via keyboard or MIDI inputs. \aria{} is an AI-powered songwriting assistant integrated into Hookpad, capable of generating melody or chord continuations and infills based on user-provided inputs. We selected these participants because Hookpad supports the earliest stages of songwriting, offering insight into the ideation process, while \aria{} allows us to examine how users already incorporate contextual generation into their creative workflows.

All participants identified themselves as hobbyists, with songwriting years ranging from 1.5 years to 5 years (M=3.38, SD=1.19). 
These participants engaged in various genres of music creation: three participants in pop songs and others in rock, bluegrass, punk, classical, and jazz. 
Participants reported engaging in songwriting activities at least once a week and using \aria{} at least once when writing songs using Hookpad.

During the interview, we asked participants about sources of inspiration and how they translate these inspirations into songs. We also asked their experiences and opinions on songwriting tools they have used (including \aria{}) and desired AI functionalities to support songwriting. Participants were compensated with \$25 Amazon gift card for the 1-hour interview. The interviews were conducted online via Zoom in a semi-structured format and were recorded, manually transcribed, and coded through a thematic analysis.

\subsection{Findings}
\label{section:formative:findings}

\subsubsection{\textbf{Inspirations from diverse modalities}}
\label{section:formative:findings:modalities}
All of the songwriters mentioned drawing inspiration from various sources, which can be grouped into three categories: audio, narrative, and visual.
Seven out of eight songwriters (P1-3, P5-8) said they \textbf{often get ideas from listening to music}. P3 noted, ``\textit{I enjoy studying songs that are already proven to be great. I listen to them, take elements I like, and add my own twists.}''
Three songwriters (P4, P7-8) mentioned \textbf{drawing inspiration from narratives}, such as stories and personal experiences. P4 explained, ``\textit{I always start with a lyrical concept when writing a song. For example, I sometimes randomly turn to a page in a novel I’m reading and use that as the theme for the song.}'' Similarly, P7 found inspiration in personal experiences: ``\textit{I remember when my dog found a baby bird in a bush. It was a tense moment because we didn’t know if my dog would attack the helpless bird. I connected that experience to a song I wrote in the harmonic minor scale.}'' 
Lastly, two (P2, P8) mentioned being \textbf{inspired by visuals}, such as game scenes or movie aesthetics. P8 noted, ``\textit{Watching a movie, the aesthetic can make me want to capture that in a song. For example, an action scene might inspire me to create something up-tempo or intense.}''

\subsubsection{\textbf{Transforming multimodal inspirations into musical elements}}
\label{section:formative:findings:transform}
Once inspired, songwriters sought to translate their abstract sources of inspiration into concrete musical elements, such as chord progressions or melodies, to build upon. Seven out of eight participants (P2-8) mentioned \textbf{improvising with instruments or their voices} until they found something they liked. For instance, P4 explained, ``\textit{I have a theme, that is, nostalgic memories triggered by scent. I would sing random melodies or try out many different chord combinations until I found something that matched my song concept.}'' The process of improvisation was not always linear; sometimes, participants improvised first and connected their ideas to their initial inspiration. P7 noted, ``\textit{I have a Google doc where I dump all my lyrical concepts, but I don't always start from there. Sometimes, I'll improvise on my guitar, come up with chord progressions I like, and begin writing. Later, I often find myself connecting the song to concepts from my list.}''

Another common approach was \textbf{analyzing music through listening sessions.} Four participants (P1, P3, P5-6) mentioned analyzing either existing songs or recordings of their own improvisations. P6 described, ``\textit{I'll go to YouTube and listen to popular music. I'll note how sharps, flats, and chord progressions are used, to find what sounds pleasing to me. Then I'll transcribe them into Hookpad and expand on it to create something unique.}''

\subsubsection{\textbf{Seamless incorporation of modular AI suggestions}}
\label{section:formative:findings:incorporation}
While all participants emphatically expressed the central importance of the manual aspects of the songwriting process, they also \textbf{recognized AI tools as inspirational sources}, as mentioned by seven participants (P2-8). AI suggestions were viewed as a ``\textit{great starting point}'' (P4, P8) or useful for overcoming ``\textit{writer's block}'' (P2, P3), as P2 remarked, ``\textit{It's like having someone next to me, suggesting the next two bars.}'' 
Participants \textbf{appreciated AI tools like \aria{} for offering modular suggestions that can be easily integrated into the songwriting process}, particularly in preserving creativity, agency, and ownership in the process (P2-4). 
For example, participants described modular suggestions in MIDI format as helpful because ``\textit{you have control over how and where to use, and you can edit the suggestions}'' (P4). Generating more than that, such as an entire audio track, was seen as intrusive, as it ``\textit{takes the joy of songwriting from me}'' (P2). 
Similarly, P3 described these modular suggestions as ``\textit{sparks that promote curiosity and creativity,}'' while generating too much ``\textit{dampens my creativity.}''
In addition, concerns about integrity were raised regarding audio track generations; two (P2-3) expressed fears that AI-generated audio might be ``\textit{potentially copied from copyrighted music}'' (P3). 

However, many participants found \textbf{\aria{}'s suggestions to be lacking context for their specific inspirations} (P1-5, P7-8). P4 explained, ``\textit{I wish it would generate melodies that better fit my lyrical ideas. Most of the time, \aria{}’s outputs don’t align with my song’s theme.}''
One participant (P5) used the text-to-music model, Udio~\citep{udio}, to address this but found it tedious, as it generated full audio track that could not be easily integrated into songwriting interfaces like Hookpad or digital audio workstation (DAW): ``\textit{I found Udio really good at finding harmonies for specific styles or moods I'm looking for. I generated audio with a prompt `country pop, uplifting, catchy anthem, introspective mood, love' and took the chord progression from there. But extracting the usable elements was a lot of work.}''

\subsection{Design Goals}
% \rev{We distill insights from the formative interviews into design goals for a multimodal songwriting system. We additionally take high-level inspiration from principles of user interface design~\citep{} and human-centered AI~\cite{}, as we aim to design a human-AI collaborative system.}
\rev{Guided by both our formative interviews and prior literature, we establish three Design Goals for our system:}
% Based on insights from the formative interviews, we formalize the following Design Goals for our system:

\begin{enumerate}
    \item[\textbf{DG1:}] \textbf{Incorporate multimodal inspirations.} 
    An AI songwriting system can support creative processes by enabling users to incorporate inspirations from diverse sources into their songwriting process. Based on the formative study, sources that users draw inspiration from are visual, narrative, and existing music.
    
    \item[\textbf{DG2:}] \textbf{Transform abstract ideas into musical elements.} 
    \rev{The system should support transforming high-level, abstract ideas into concrete musical elements. 
    As informed by prior studies, 
    this transformation process 
    should ensure 
    user control over iterative generation of musical elements~\citep{shneiderman2022human, 10.5555/3033040}, 
    while aligning outputs with users' creative intentions~\citep{10.1145/302979.303030}.}
         
    \item[\textbf{DG3:}] \textbf{Provide modular musical building blocks.} 
    \rev{Consistent with prior studies~\citep{10.1145/302979.303030, Resnick2018}, participants in our formative study 
    preferred modular, editable outputs that can be seamlessly integrated into their workflow. 
    As such, the system should focus on generating modular musical elements---such as chord progressions---that can be refined and directly integrated into songwriting interfaces (e.g., DAWs). 
    }
    
\end{enumerate}

\section{\system{}: Overview}
\label{section:amuse}

\begin{figure*}[t!]
    \centering
    \includegraphics[width=\linewidth]{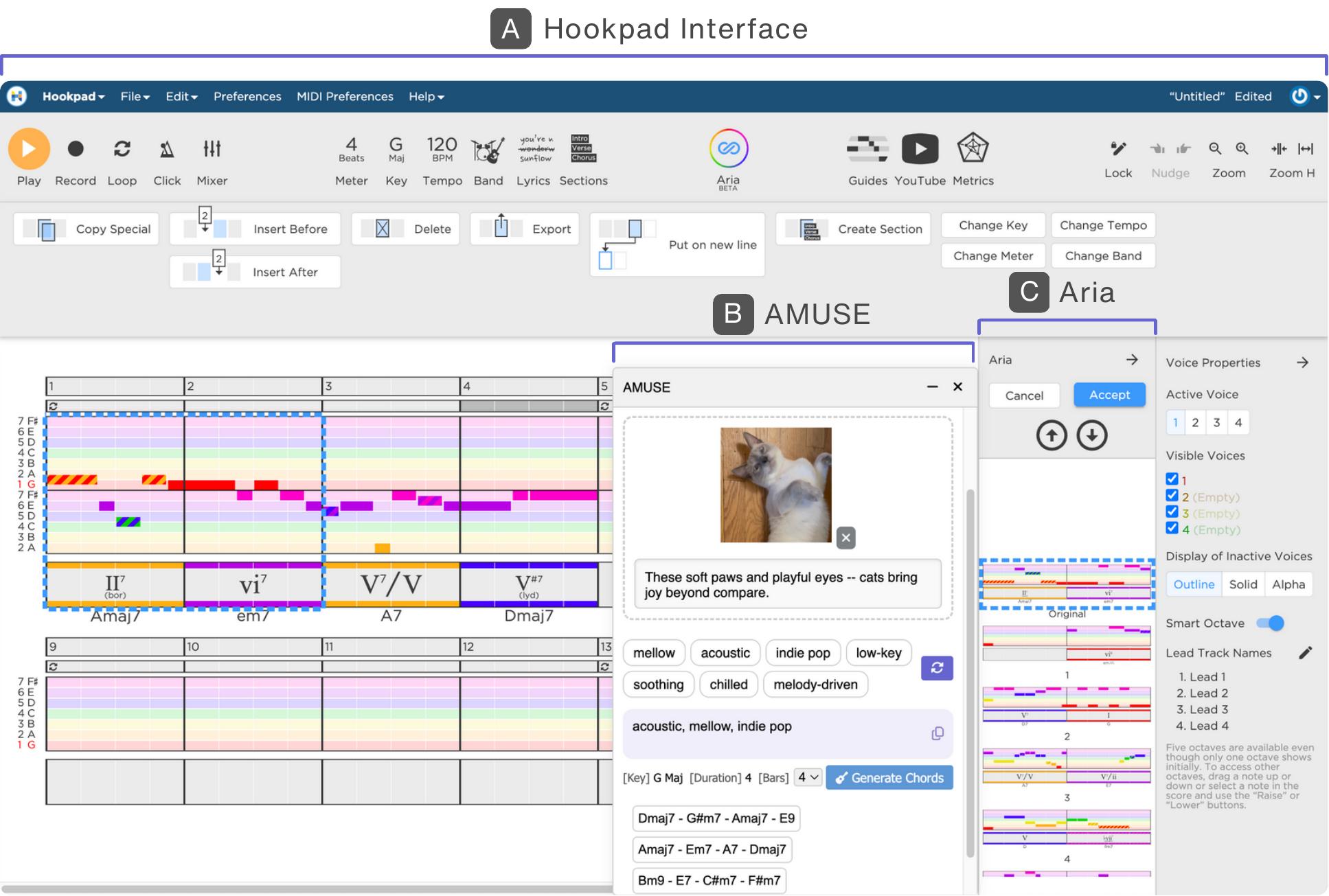}
    \caption{
    Screenshot of the songwriting interface used in the user study. The main workspace is the Hookpad interface (A), where users can input melodies and chords using either the keyboard or MIDI devices. \system{} (B), a Chrome extension, appears as a floating window within the Hookpad interface. Users can freely open, close, move, and resize this window. The current view of \system{} displays the Chord Generator, which generates chord progressions from user-provided images or text. Participants also used \aria{} (C), a tool for generating melodies and chords based on the content already written in the Hookpad interface.
    }
    \Description{A screenshot of the songwriting interface used in the user study. The Hookpad Interface (A) serves as the main workspace where users can input melodies and chords using a keyboard or MIDI devices. Amuse (B) is a Chrome extension that appears as a floating window within the Hookpad interface, displaying the Chord Generator based on user-provided images or text. Aria (C) is another tool used to generate melodies and chords based on the content already written in the Hookpad interface.}
    \label{fig:systemoverview}
\end{figure*}

% AMUSE: *A*I-powered *Mu*ltimodal *S*ongwriting Support Syst*e*m
Based on our design goals, we present \system{}~(Figure~\ref{fig:systemoverview}), a songwriting assistant that transforms high-level multimodal inspirations into chord progressions. \system{} supports diverse sources of inspiration, including text, images, and audio (DG1), and translates these abstract inputs into coherent chord progressions (DG2) by either keyword-guided chord generation or audio transcription. By focusing on chord progressions as the system outputs (DG3), \system{} provides users with fundamental components that can be seamlessly integrated into their compositions within the songwriting interface. 
In this section, we first provide an overview of \system{} (\S\ref{section:amuse:overview}), explain the interactions and technical details within \system{}'s interface (\S\ref{section:amuse:chordgenerator}-\S\ref{section:amuse:chordtranscriber}), and implementation (\S\ref{section:amuse:implementation}).

\subsection{System Overview}
\label{section:amuse:overview}

\system{} is built as a Chrome extension that works on top of Hookpad~\cite{hookpad}, a widely-used online songwriting tool developed by Hooktheory~\cite{hooktheory} (see Figure~\ref{fig:systemoverview}). The decision to develop \system{} as an 
%extension 
companion tool to Hookpad 
rather than a standalone application was driven by the robust songwriting functionalities already present in Hookpad, such as MIDI input, playback, and instrumentation options, making it an efficient platform to build upon without the need to duplicate existing functionalities. 
Furthermore, Hookpad includes \aria~\cite{aria, anticipatory}, an AI songwriting assistant that generates melody/chord continuations or 
%infillations 
infills 
based on user-provided melodies and chords. \aria{} 
%represents traditional 
is representative of an emerging family of contextual AI music assistants~\cite{cococo, bachdoodle, expressive}, i.e.,~ones that generate musical outputs based on existing musical material.
%generating low-level musical outputs based on existing low-level musical inputs. 
In contrast, \system{} is designed to facilitate 
%high-to-low-level idea 
multimodal 
transformations, 
%where high-level multimodal inspirations are translated into low-level musical ideas. 
complementary to the contextual inputs to \aria{}. 
By integrating \system{} with Hookpad and \aria{}, we create a unique environment to study the interaction between these different forms of music AI assistance. 
This setup not only allows us to explore the synergies that \system{} brings to the previous songwriting interfaces but also provides valuable insights into their collaborative potential in enhancing the songwriting process.

\system{} comprises two main features: Chord Generator (\S\ref{section:amuse:chordgenerator}) and Chord Transcriber (\S\ref{section:amuse:chordtranscriber}). 
Chord Generator processes text and image inputs, extracting relevant musical keywords that users can select and refine. These keywords are then used to guide the generation of chord progressions that align with the mood and concepts embodied by the keywords. 
On the other hand, Chord Transcriber accepts audio inputs and transcribes them into chord progressions.
When users select (click) chords suggested by \system{}, they are instantly integrated into the Hookpad interface, where they can be played back and further refined. Users can change instrumentation, timbre, and play the sounds within the Hookpad interface, and continue songwriting based on \system{}'s suggestions.

% \subsection{Interface}
% \label{section:amuse:interface}

\subsection{Chord Generator}
\label{section:amuse:chordgenerator}
\begin{figure*}[t!]
    \centering
    \includegraphics[width=\linewidth]{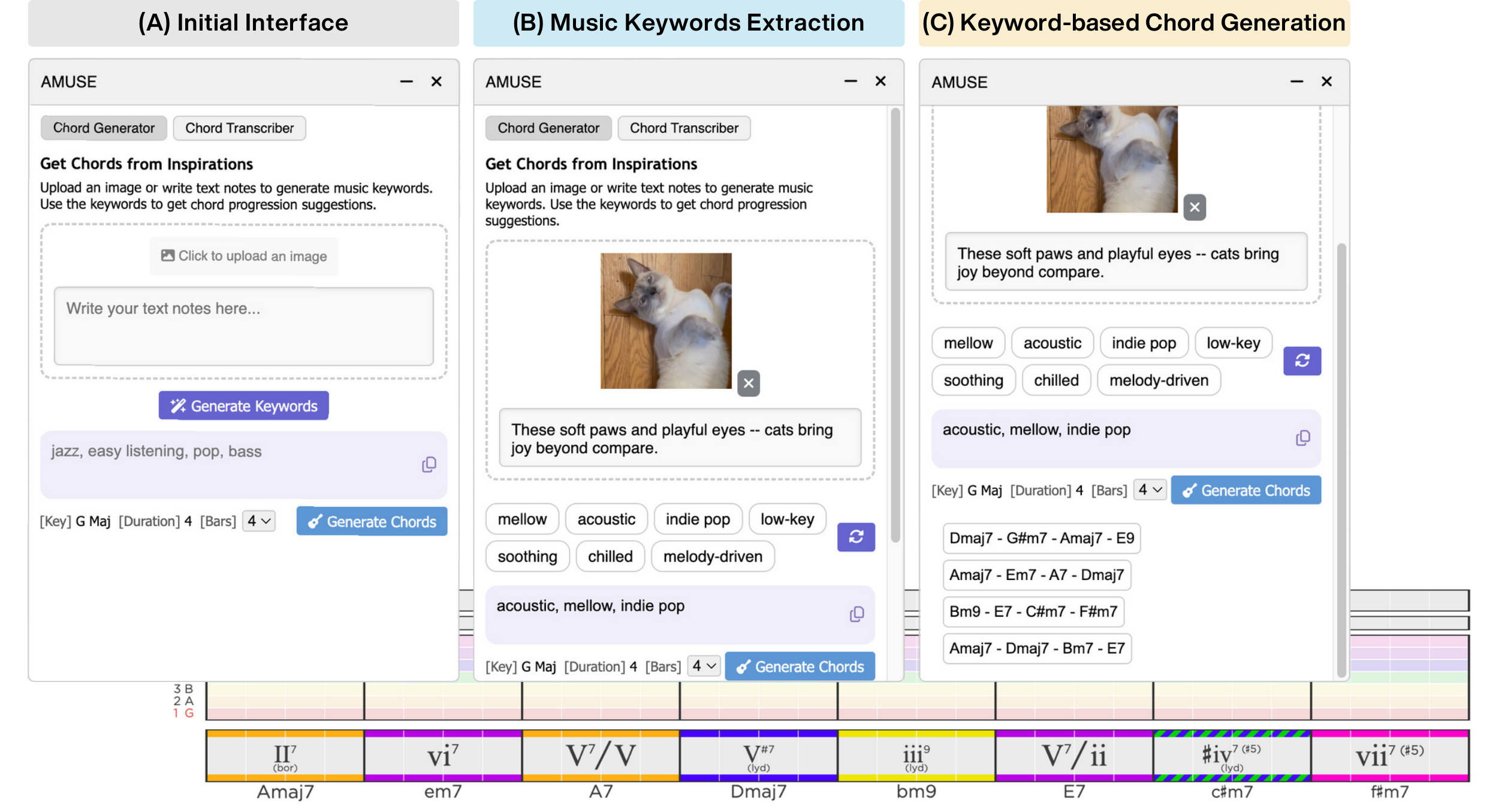}
    \caption{
    Overview of Chord Generator in \system{}. (A) Initial Interface: Users can upload an image or type text, which are used to generate music keywords. Users can also directly write music keywords in the keyword editor.
    (B) Keyword Extraction: Upon clicking the ``Generate Keywords'' button, \system{} suggests music keywords based on the multimodal inputs. 
    User-selected keywords are automatically pasted into the keyword editor (`acoustic, `mellow,' and `indie-pop' in the figure). 
    (C) Keyword-based Chord Generation: Upon clicking the `Generate Chords' button, \system{} suggests four chord progressions based on the keywords. Users can choose between 3-bar or 4-bar progressions (default: 4 bars). The key is automatically detected from the song configuration in Hookpad (G Maj in the figure). 
    Clicking a chord progression automatically pastes it into Hookpad (in the figure, `Amaj7-Em7-A7-Dmaj7' and `Bm9-E7-C\#m7-F\#m7' are selected), where users can play the audio and make further edits. 
    }
    \Description{
    Illustration of Chord Generator workflow and interface. (A) Users can upload an image or type text to generate music keywords. Alternatively, they can also directly input keywords into the editor to generate chord suggestions. (B) Upon uploading an image or text, AMUSE generates music keywords based on the input. In this example, keywords like "mellow," "acoustic," and "indie pop" are extracted from the provided image and text. (C) After selecting keywords, clicking the "Generate Chords" button generates chord progressions. Users can choose from 3-bar or 4-bar progressions, and the detected key is automatically matched to the song configuration in Hookpad.
    }
    \label{fig:chordgenerator}
\end{figure*}

\subsubsection{Chord Generator Interface.}
The Chord Generator interface is illustrated in Figure~\ref{fig:chordgenerator}. 
Upon opening the Chord Generator, users are prompted to provide an image, text input, or both, which are then used to extract relevant music keywords (Figure~\ref{fig:chordgenerator}A). 
When users click the `Generate Keywords' button, the system suggests keywords based on the multimodal inputs (Figure~\ref{fig:chordgenerator}B). 
Users can select from the suggested keywords, with selected keywords automatically inserted into the keyword editor for further refinement.
After finalizing the keywords, users can click the `Generate Chords' button to receive chord progressions that align with the selected music keywords (Figure~\ref{fig:chordgenerator}C). The interface reads the key configuration (e.g., G Maj) in the Hookpad editor interface and generates the chord progressions in the specified key. Users can also specify the number of bars (3 or 4) for the chord progressions.
Upon clicking the button, \system{} generates and displays four chord progressions, drawing on prior research that demonstrates multiple suggestions enhance creativity and are preferred in ideation tasks~\citep{dsiiwa, multipleparallel, dang2023choice, cococo}. We specifically provide four suggestions per query, following previous work on human-AI collaborative music creation system~\citep{cococo}. 
By clicking on any of the suggestions, the chosen progression is automatically inserted into the Hookpad editor, where users can further modify or play the chords to integrate them into their compositions.

\subsubsection{Keyword-based Chord Generation Pipeline.}
We adopted a keyword-based interaction framework between multimodal inputs and chord progressions to help users create meaningful semantic connections between two modalities that are otherwise difficult to link due to their abstract and morphologically different nature~\citep{elephant}.
Keywords serve as an intermediary communication layer, inspired by practices in creativity support tools in the vision domain~\citep{opal, 3dalle, 10.1145/3613904.3642794}. To extract music-related keywords that align with user inputs, we leverage the multimodal LLM, GPT-4o~\citep{gpt4o}, instructing it with a list of music styles and genre-specific keywords crawled from a music keyword wiki~\citep{sunowiki}. The prompt used for keyword generation can be found in Appendix~\ref{appendix:prompts}.

For keyword-based chord generation, to ensure that the generated chords are diverse, musically coherent, and relevant to input keywords, we developed a rejection-sampling-based chord generation method that integrates an LLM with a language model prior trained over chord data. The detailed algorithm and its implementation are discussed in \S\ref{section:amuse:generation}.

\begin{figure*}[t!]
    \centering
    \includegraphics[width=\linewidth]{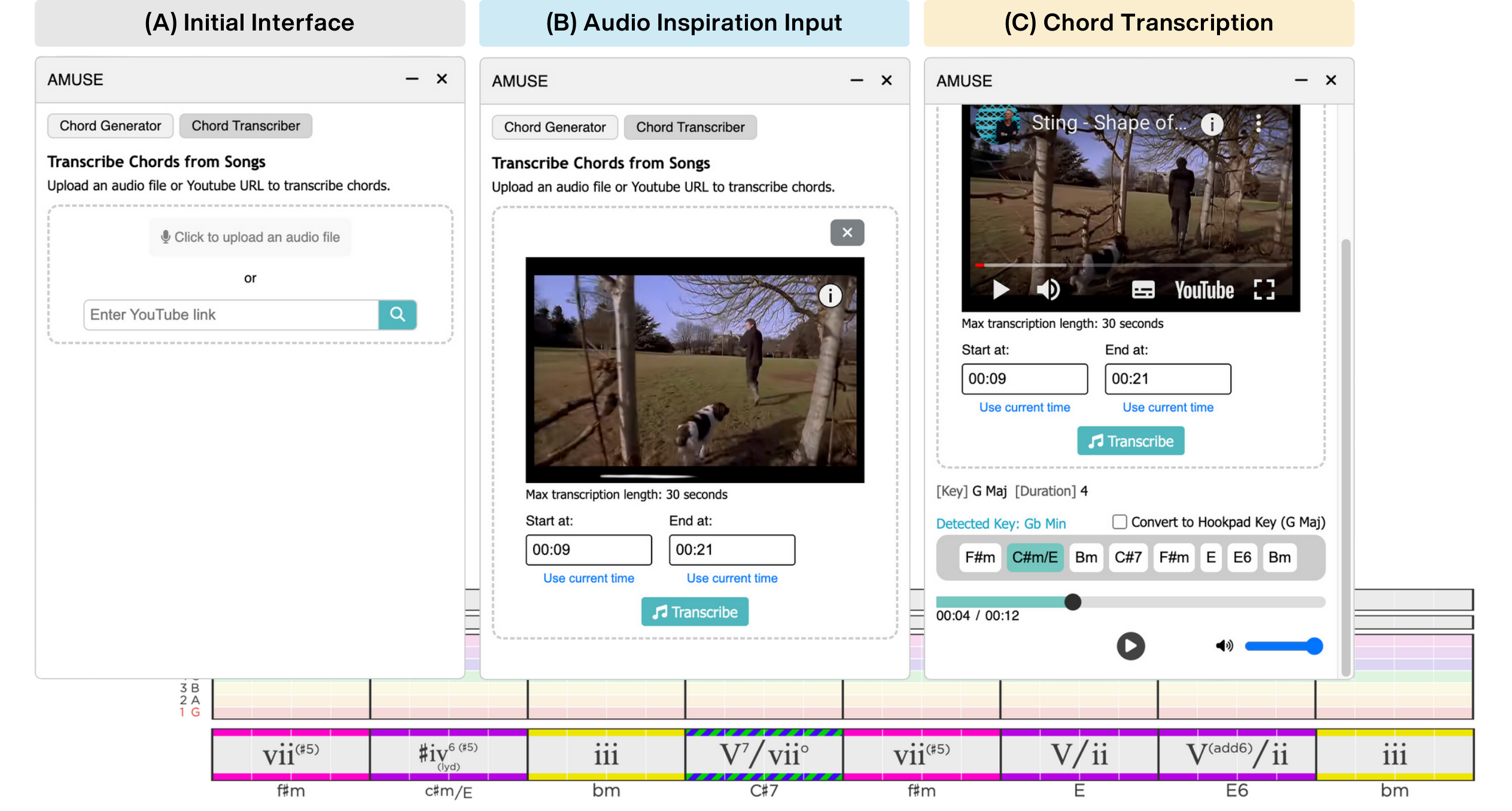}
    \caption{
    Overview of Chord Transcriber in \system{}. (A) Initial Interface: Users can upload a local audio file or enter a YouTube URL.
    (B) Audio Inspiration Input: With an audio preview, users can select the desired segment for transcription by specifying start and end times (maximum 30 seconds). 
    (C) Chord Transcription: \system{} detects the key (shown as Gb Min) and chords of the selected audio segment. 
    Users can play the audio segment and the chord is highlighted in sync with the playback (in this figure, C\#m/E). 
    Clicking on a chord pastes it into the Hookpad interface. If the ``Convert to Hookpad Key'' option is checked, the chords are transposed to match the key the user is working on (G Maj in the figure).
    }
    \Description{
    Illustration of Chord Transcriber workflow and interface. (A) Users can either upload an audio file or enter Youtube link. (B) After inputting the audio, users select the start and end times for transcription (up to 30 seconds), and AMUSE shows the preview of the audio. (C) AMUSE detects the key and transcribes the chords. The detected keys are in the gray colored box, where clicking it will paste it into the Hookpad interface. The detected chords can be played back in sync with the audio and automatically transposed to the user’s current key if the "Convert to Hookpad Key" option is selected.
    }
    \label{fig:chordtranscriber}
\end{figure*}

\subsection{Chord Transcriber}
\label{section:amuse:chordtranscriber}

The Chord Transcriber interface is illustrated in Figure~\ref{fig:chordtranscriber}. 
Users can either upload a local audio file or enter a YouTube URL for transcription (Figure~\ref{fig:chordtranscriber}A). Once the audio source is uploaded, users can preview the audio and set the transcription window, with a maximum length of 30 seconds (Figure~\ref{fig:chordtranscriber}B). 
Upon clicking the `Transcribe' button, \system{} processes the request with a 10$-$15 seconds delay and displays the detected key and chords for the selected audio segment (Figure~\ref{fig:chordtranscriber}C). 
The transcribed audio can be played back with the corresponding chords highlighted in sync with the playback. Users can click on any chord to automatically paste it into the Hookpad editor for further refinement. To accommodate key differences between the detected key and the key the user is working in, the `Convert to Hookpad Key' option allows for automatic transposition of the chords to match the current working key in Hookpad. \rev{For chord transcription, we use an established chord transcription API provided by Music AI, a company that creates and provides AI platforms for music and audio production~\citep{moisesapi}.}

\subsection{Implementation}
\label{section:amuse:implementation}
We implemented the frontend of \system{} (Chrome extension) 
%using pure 
in 
JavaScript and CSS. The backend was implemented as a Flask server, and we call the OpenAI API~\cite{openaiapi} for all LLM-based functionality (multimodal inputs to keywords, keywords to chords). Regarding the LLM configurations, we use \texttt{gpt-4o-2024-05-13} and set the temperature to $1.0$ for all components to promote creativity and diversity in generations. 
\section{Keyword-Based Chord Progression Generation}
\label{section:amuse:generation}

Directly generating chord progressions conditioned on music keywords presents a challenge due to the lack of paired training data linking music keywords to chord progressions. 
To address this issue, we propose a method that combines the general capabilities of LLMs and a unimodal chord generation model specifically trained on chord data. 
We first introduce the generation goals we aim to achieve (\S\ref{section:amuse:generation:goal}). We then detail the components of our pipeline designed to address these goals, including a prompting technique for diverse outputs (\S\ref{section:amuse:generation:diverse}) and a rejection sampling methods for contextually coherent outputs (\S\ref{section:amuse:generation:contextual}). Finally, we describe the implementation of our chord progression generation method deployed in \system{} (\S\ref{section:amuse:generation:implementation}).

\subsection{Generation Goals}
\label{section:amuse:generation:goal}

When generating chord progressions from user-provided keywords, we consider three desiderata: \textbf{diversity}, \textbf{relevance}, and \textbf{coherence}.
\textbf{Diversity} ensures that users receive a variety of options, as offering multiple suggestions has been shown to enhance creativity~\cite{dsiiwa} and is often preferred in ideation tasks~\cite{multipleparallel, dang2023choice, cococo}. 
\textbf{Relevance} ensures that the generated chord progressions closely align with the user-provided keywords, reflecting the musical qualities or moods the user intends to capture.
\textbf{Coherence} ensures that the chord progressions are musically coherent (i.e., follow real music data distribution). 
Ensuring both relevance and coherence is crucial as misaligned or low-quality suggestions can negatively impact user experience in creativity support tools~\cite{dsiiwa, metaphoria, 10.1145/3172944.3172983, manjavacas-etal-2017-synthetic}.

Formally, let $P(\mathbf{x}| \mathbf{c})$ be a target conditional distribution of chord progression $\mathbf{x}$ capable of generating diverse, relevant, and coherent suggestions given keywords control $\mathbf{c}$. Modeling $P(\mathbf{x}| \mathbf{c})$ is limited by the scarcity of music data that contains \textit{paired} keywords and chord progressions. As a result, we aim to find a good approximation of $P(\mathbf{x}| \mathbf{c})$ by using an \textit{unpaired} chord progression dataset $\mathcal{D} = \{ \mathbf{x}_i \}_{i=1}^{N}$ and an LLM (e.g., GPT-4o~\citep{gpt4o}) that has a general understanding of language and general music-related context but lacks the direct knowledge of harmonic structures seen in real-world music data (as it is not explicitly trained on music data).
Specifically, we define $Q(\mathbf{x}| \mathbf{c})$ as an LLM that generates chord progressions $\mathbf{x}$ given keywords $\mathbf{c}$ as prompts. 
We select an instruction-tuned LLM (GPT-4o~\cite{gpt4o}) for $Q(\mathbf{x}|\mathbf{c})$ due to its enhanced ability to follow user instructions that are critical for generating contextually relevant chord progressions from music keywords. 
\rev{These models are pretrained on large-scale text data from public webpages spanning diverse topics~\citep{gpt4osystemcard}, which include music-related content (for instance, chord transcriptions~\citep{srcexample2} or music theory blogs~\citep{srcexample1}). As such, we assume the model has a general understanding of music and can infer relationships between musical elements (e.g., chords) and high-level descriptions (e.g., mood or style).}
% These models are fine-tuned on diverse text datasets, which likely include text related to music, allowing them to infer relationships between musical elements (e.g., chords) and high-level descriptions (e.g., mood or style)~\citep{gpt4osystemcard}.
While $Q(\mathbf{x}| \mathbf{c})$ is capable of keyword-controlled generation, we find this leads to a very sharp distribution (i.e., lacks diversity; we later show in our experiments that its outputs exhibit low diversity).
We mitigate this lack of diversity by introducing a prompting technique for LLMs to generate diverse samples (\S\ref{section:amuse:generation:diverse}). 
We also find the distribution $Q(\mathbf{x}| \mathbf{c})$ inherently has low coherence (i.e., deviates from real music distribution) as LLMs are not explicitly trained on music data. We thus present a rejection sampling strategy using a unimodal chord generation model $P(\mathbf{x})$ to select coherent outputs (\S\ref{section:amuse:generation:contextual}). 
Based on these techniques, we describe the implementation of our chord progression generation process deployed in \system{} (\S\ref{section:amuse:generation:implementation}).

\subsection{Prompting Technique For Diverse Chord Progressions}
\label{section:amuse:generation:diverse}

In creative tasks, generating diverse suggestions with LLMs typically involves multiple queries to the model~\cite{coauthor, multipleparallel, dang2023choice}. However, instruction-tuned LLMs such as InstructGPT~\citep{instructgpt} and GPT-4o~\citep{gpt4o} are prone to a homogenization effect, where multiple queries often yield repetitive responses~\citep{padmakumar2024does, anderson2024homogenization, doshi2024generative, zhang2024forcing}. This is particularly problematic in our context, where generating diverse outputs is crucial for fostering creativity.

To address this challenge of generating diverse chord progressions with instruction-tuned LLMs, we employ a prompting technique that instructs an LLM to generate a large collection of chord progressions. Our prompt takes user-written keywords, key (e.g., C maj) and bar counts (e.g., 4 for 4-bar chord progression) as inputs. The prompt then instructs the model $Q(\mathbf{x}|\mathbf{c})$ to generate $N$ diverse chord progressions that reflect the mood and concept of the user keywords. To enhance diversity, the prompt contains instructions to vary chord components (e.g., root, quality, extensions, alterations), progression patterns (e.g., diatonic, chromatic), and cadences. The full prompt is detailed in Appendix~\ref{appendix:prompts:chordsamuse}.

\subsection{Chord Progression Generation By Rejection Sampling}
\label{section:amuse:generation:contextual}

Given $N$ diverse chord progressions generated by $Q(\mathbf{x}|\mathbf{c})$ (\S\ref{section:amuse:generation:diverse}), the most naive approach would be randomly sampling four progressions and showing them to the users. However, this is not the best approach as $Q(\mathbf{x}|\mathbf{c})$ generations could be less coherent, i.e., do not align with the statistical properties of chord progressions in real music data, as these language models are not explicitly trained on music data~\citep{gpt4osystemcard}.
Consequently, relying solely on $Q(\mathbf{x}|\mathbf{c})$ for chord generation may produce noisy or less conventional progressions that do not meet the musical expectations of users.
Therefore, our aim is to filter and select from the generations that are musically coherent, i.e., following real-world music data distribution. To achieve this, we employ rejection sampling~\cite{rejectionsampling}, a Monte Carlo algorithm that allows us to sample from a target distribution that may be difficult to sample from directly (i.e., $P(\mathbf{x}|\mathbf{c})$), by using a proposal distribution that is easier to sample from (i.e., $Q(\mathbf{x}|\mathbf{c})$).

\paragraph{\textbf{Method.}}

We perform rejection sampling~\citep{rejectionsampling} to improve the coherence of $Q(\mathbf{x}|\mathbf{c})$:
\begin{align*}
    u\sim \text{Unif}(0,1),\,\,\text{{accept} {\bf if }} u < \frac{P(\mathbf{x} | \mathbf{c})}{M \cdot Q(\mathbf{x} | \mathbf{c})}\,\, \text{{\bf else} reject},
\end{align*}
\noindent where $M>0$ is a constant that ensures the inequality holds for all samples and scales the acceptance ratio. $M$ is chosen such that $M \geq \max_{\mathbf{x}} \left( \frac{P(\mathbf{x} | \mathbf{c})}{Q(\mathbf{x} | \mathbf{c})} \right)$, ensuring that the acceptance rate remains valid across the entire sampling process.

Since we do not know ground-truth $P(\mathbf{x} | \mathbf{c})$, as well as the value of $Q(\mathbf{x} | \mathbf{c})$ (we can only sample from this distribution), we consider the following alternative using Bayes Rule: 
\begin{align*}
    \frac{P(\mathbf{x} | \mathbf{c})}{M \cdot Q(\mathbf{x} | \mathbf{c})}
     = 
    \frac{P(\mathbf{c}|\mathbf{x})P(\mathbf{x})Q(\mathbf{c})}{M \cdot Q(\mathbf{c} | \mathbf{x})Q(\mathbf{x})P(\mathbf{c})} = \frac{P(\mathbf{c}|\mathbf{x})P(\mathbf{x})}{M \cdot Q(\mathbf{c} | \mathbf{x})Q(\mathbf{x})} \approx \frac{P(\mathbf{x})}{M \cdot Q(\mathbf{x})},
\end{align*}

\noindent since $P(\mathbf{c})=Q(\mathbf{c})$ as they are the same predefined keyword distributions and we assume $P(\mathbf{c}|\mathbf{x}) \approx Q(\mathbf{c} | \mathbf{x})$ as the generated progression $\mathbf{x}$ from LLMs closely align with $\mathbf{c}$. Thus, we perform rejection sampling by calculating the ratio $\frac{P(\mathbf{x})}{M \cdot Q(\mathbf{x})}$, where both $P(\mathbf{x}$) and $Q(\mathbf{x})$ can be learned by training two deep neural networks with the dataset $\mathcal{D}$ and a large collection of generated chord from LLMs, respectively.

\begin{algorithm}[t!]
\caption{Chord Progression Generation Pipeline in \system{}}
\label{algo:generation}
\begin{algorithmic}[1]
\State \textbf{Input:} Keywords $\mathbf{c}$ %, scaling factor $M$, desired number of accepted samples $n$ %target model $\tilde p$, proposal model $\tilde q$, 
\State \textbf{Output:} Accepted chord progressions $\mathbf{Y}$
\State $\mathbf{X}\gets$ Candidate chord progressions $\{\mathbf{x}_i\}_{i=1}^{N}$ from $Q(\mathbf{x}|\mathbf{c})$
\State $\mathbf{Y}\gets$ $\emptyset$
\For {$\mathbf{x}$ in $\mathbf{X}$} 
    \State Sample $u \sim \text{Unif}(0,1)$ 
    \If {$u < \frac{P(\mathbf{x})}{M Q(\mathbf{x})}$} \Comment{Compute acceptance probability}
        \State $\mathbf{Y} \gets \mathbf{Y} \cup \{\mathbf{x}\}$ \Comment{Accept the generated chord $\mathbf{x}_i$}
        %Add $\mathbf{x}_i$ to $\mathbf{X}$
    \EndIf
\EndFor
\If {$|\mathbf{Y}| < 4$}
    \State $\mathbf{Y} \gets \mathbf{Y} \cup \{\text{top-}k(\mathbf{x}_i : \frac{P(\mathbf{x}_i)}{M Q(\mathbf{x}_i)})\}$, where $k = 4 - |\mathbf{Y}|$ 
    \par
    \Comment{Add top-$k$ samples if $\mathbf{Y}$ is smaller than the target size}
\EndIf
\State \textbf{Return} $\mathbf{Y}$
\end{algorithmic}
\end{algorithm}

\paragraph{\textbf{Implementation.}}

For $P(\mathbf{x})$ and $Q(\mathbf{x})$, we train two Long Short-Term Memory~(LSTM) models. 
Specifically, for $P(\mathbf{x})$, we use the HookTheory dataset~\cite{sheetsage}, which comprises 50 hours of melody and chord progression annotations derived from Hooktheory's TheoryTab database~\cite{theorytab}. This dataset spans a wide range of genres, including pop, rock, EDM, jazz, and classical. For training $Q(\mathbf{x})$, we use chord progressions generated by GPT-4o~\citep{gpt4o}. We use the same prompt used for generating diverse chord suggestions (\S\ref{section:amuse:generation:diverse}). To marginalize the influence of specific keywords during generation \rev{(i.e., generalize across diverse keywords)}, we randomly sample keywords from the music keyword wiki~\cite{sunowiki} for each generation. This process resulted in a total of 25,000 4-bar chord progressions with keywords abstracted into the underlying distribution. 
After fitting these models, the constant \( M \) in the rejection sampling algorithm was determined by calculating the ratio \( \frac{P(\mathbf{x})}{Q(\mathbf{x})} \) over all chord progressions generated by GPT-4o. We selected the 95th percentile value to avoid the influence of extreme outliers, resulting in $M=7.64$. We provide training details and hyperparameter values in Appendix~\ref{appendix:techdetails}.

\subsection{Deploying Chord Progression Generation Pipeline in \system{}}
\label{section:amuse:generation:implementation}

The obtained constant $M = 7.64$ and the trained networks, $P(\mathbf{x})$ and $Q(\mathbf{x})$, are deployed on the backend server to sample LLM-generated chord progressions. To approximately match four chord progressions, we set the number of chord generations to $N=30$ in the deployment (30/$M$ $\approx$ 4). During this process, each progression generated by $Q(\mathbf{x}|\mathbf{c})$ is assessed against the acceptance threshold determined by our rejection sampling method.
To ensure we present at least four samples to users, if fewer than four samples are accepted, we select the top-k samples, $k$ = 4 - \textit{num accepted samples}. These samples are ranked by their acceptance probability, $\frac{\tilde p(\mathbf{x})}{M \tilde q(\mathbf{x})}$. The full sampling process is detailed in Algorithm~\ref{algo:generation}. 

\section{Technical Evaluation}
\label{section:techeval}

We conduct a technical evaluation of our chord progression generation approach. We assess whether our prompting technique results in \textbf{diverse} chord progressions (\S\ref{section:techeval:prompting}) as well as whether our rejection sampling can sample chord progressions that are \textbf{relevant} to keywords and musically \textbf{coherent} (\S\ref{section:techeval:rejsampling}).

\subsection{Evaluation of Prompting Technique}
\label{section:techeval:prompting}
To assess the effectiveness of our prompting technique (\S\ref{section:amuse:generation:diverse}) in generating more \textbf{diverse} chord progressions, we conduct an automatic evaluation comparing our approach against a conventional method of generating multiple suggestions~\cite{coauthor, multipleparallel, dang2023choice}.

\paragraph{\textbf{Study Setup.}}
We compare the diversity of chord progressions produced by our prompting technique with those generated by the conventional method (\textit{baseline}), which involves querying a model multiple times to generate individual suggestions. In the baseline method, a single 4-bar chord progression is generated per query and repeated 30 times, resulting in a set of 30 4-bar chord progressions. The prompt for the baseline is similar to ours but lacks instructions to generate diverse progressions in a batch (See the full prompt in Appendix~\ref{appendix:prompts:chordsbaseline}). In contrast, our technique generates all 30 4-bar chord progressions in a single batch using a single prompt. 

To quantify the diversity of chord progressions, we employ Self-BLEU~\cite{selfbleu} as our metric. 
Self-BLEU score measures the diversity of the generated data, with a higher Self-BLEU score indicating less diversity among the chord progressions. Specifically, for each set of 30 4-bar chord progressions, we compute the BLEU score~\cite{bleu} for each individual progression by treating it as the hypothesis and the remaining 29 progressions as references. The Self-BLEU score is then obtained by averaging the BLEU scores across all 30 progressions in the set. 
This process is repeated across 100 distinct pairs of sets (i.e., 100 sets of 30 4-bar chord progressions generated by our technique and 100 corresponding sets generated by the baseline method). All chord progressions are generated in C, and the music keywords are randomly sampled from the music keyword wiki~\cite{sunowiki} for each generation. We report the average Self-BLEU score across these 100 pairs for each condition.

\paragraph{\textbf{Results.}} 

\begin{table*}[t!]
    \centering%\small
    \begin{minipage}{.3\linewidth}
    \centering
    \caption{\system{} offers more diverse chord suggestions than the baseline by instructing an LLM to generate diverse chords in a batch, as indicated by the Self-BLEU~\cite{selfbleu} scores for chord progression generations. $\downarrow$ indicates lower values are better. Each value represents the Mean$\pm$Std score across 100 chord progression sets, where each set contains 30 chord progressions.}
    \begin{tabular}{l c}
    \toprule
    Method & Self-BLEU~$\downarrow$ \\
    \midrule
    Baseline  &  0.61{\scriptsize$\pm$0.18} \\
    \rowcolor{aliceblue} \system{} &  0.30{\scriptsize$\pm$0.12} \\
    \bottomrule
    \end{tabular}
    \label{tab:eval:selfbleu}
    \end{minipage}
    \hspace{1.5em}
    \begin{minipage}{.6\linewidth}
    \centering
    \caption{\system{} generates keyword-conditioned chord progressions that are more coherent, i.e., closer to real music data distributions, as indicated by Jensen-Shannon Divergence~(JSD)~\cite{jsd} between each chord generation method and music data~\cite{sheetsage}. $\downarrow$ indicates that lower values are better. We compute JSD for unigram/bigram chord distributions. LSTM Prior refers to chords generated by an LSTM trained on music data ($P(\mathbf{x})$). GPT-4o refers to chords generated by GPT-4o ($Q(\mathbf{x}|\mathbf{c})$) with keywords marginalized. \system{} represents rejection-sampled GPT-4o using LSTM Prior.}
    \begin{tabular}{l c c c}
        \toprule
         Method & \begin{tabular}[c]{@{}c@{}}Keyword \\ Conditional \end{tabular} & Unigram~$\downarrow$ & Bigram~$\downarrow$  \\
         \midrule
         LSTM Prior~(N=25,000) & No & 0.15 & 0.30 \\
         GPT-4o~(N=25,000) & Yes & 0.42 & 0.57 \\
         \rowcolor{aliceblue} \system{}~(N=3,242) & Yes & 0.27 & 0.46 \\
         \bottomrule
        % \toprule
        % Q(X): unigram 0.37, bigram 0.53
    \end{tabular}
    \label{tab:eval:jsd}
    \end{minipage}
\end{table*}

Table~\ref{tab:eval:selfbleu} shows the result for the diversity. On average, our prompting technique (\system{}) results in a lower Self-BLEU score compared to the conventional method (Baseline). This indicates that the chord progressions generated by our technique are more diverse than those generated by the baseline method. 

\subsection{Evaluation of Rejection Sampling Approach}
\label{section:techeval:rejsampling}

We evaluate the \textbf{coherence} and \textbf{relevance} of sampled chord progressions generated through our rejection sampling approach (\S\ref{section:amuse:generation:contextual}). Specifically, our goal is to assess whether the resulting progressions are (i)~musically coherent and (ii)~relevant to keywords. 
We first run an automatic evaluation to compare the distribution of the sampled progressions with that of a real music dataset~\cite{hooktheory}, where we interpret closer match as greater musical coherence. 
We then perform a listening study with musicians to assess both the musical coherence and keyword relevance of the progressions.

\subsubsection{\textbf{Automatic Evaluation}}
\label{section:techeval:rejsampling:automatic}

\paragraph{\textbf{Study Setup.}}
The aim of rejection sampling is to align LLM-generated chord progressions with real music data distribution. We quantitatively assess this alignment by computing the Jensen-Shannon Divergence~(JSD)~\cite{jsd} between the distributions of the generated chord progressions and real music data. 
JSD measures the similarity between two probability distributions, with values ranging from 0 (perfect similarity) to 1 (complete divergence). Intuitively, it captures how much the chord distributions from the model differ from those found in real music, offering insight into how well the generated chord progressions reflect real-world harmonic patterns. 
Specifically, we compare unigram and bigram distributions of chord tokens and use them to calculate JSDs. 
We use the HookTheory dataset~\cite{sheetsage} as the real music dataset. We pre-process the chords and get $N=25,601$ chord progressions, all transposed into C. 
For generating chord progressions with GPT-4o, we randomly sampled music keywords from the music keyword wiki~\citep{sunowiki} to marginalize the effect of specific keyword choices. 
We calculate the JSD between the real music data and three different chord generation conditions:

\begin{enumerate}
    \item LSTM Prior ($N$=25,000): Chord progressions generated by an LSTM model trained on the HookTheory data ($P(\mathbf{x})$).
    \item GPT-4o ($N$=25,000): Chord progressions generated by GPT-4o ($Q(\mathbf{x}|\mathbf{c})$) with keyword marginalized (i.e., prompted with random keywords).
    \item \system{} ($N$=3,242): Rejection-sampled GPT-4o using $P(\mathbf{x})$ (LSTM Prior) and $Q(\mathbf{x}).$
\end{enumerate}

\paragraph{\textbf{Results.}} Table~\ref{tab:eval:jsd} presents the alignment of each distribution with real music data, as measured by JSD. The results indicate that applying rejection sampling improves the alignment of the chord progressions generated by GPT-4o with the real music data distribution from the Hooktheory dataset. While the initial LLM-generated chord progressions (GPT-4o) exhibited notable divergence from real-world chord patterns, the use of rejection sampling reduced this discrepancy.

\subsubsection{\textbf{Human Evaluation: Listening Study}}
\label{section:techeval:rejsampling:human}

\paragraph{\textbf{Study Setup.}}

We conducted a listening study to further evaluate the qualitative performance of our rejection sampling approach against two baselines. We evaluated two key aspects of the generated chord progressions: (i) musical coherence and (ii) keyword relevance. 
Following the prior work~\cite{singsong}, listeners were presented with pairs of 7-10s chord progression audio clips, each generated by different methods (ours or baselines). 

The study comprised two types of tasks: First, participants listened to a pair of 7s 4-bar chord progressions played with piano sounds and were asked to indicate which of the two ``sounds more pleasant and natural'' (musical coherence). Second, participants were presented with a set of keywords and a pair of 7-10s, 4-bar chord progressions, with instrumentation and BPM adjusted to align with the keyword characteristics. They were then asked to indicate which of the two ``better reflects the mood or concept of the given keywords'' (keyword relevance).

We generated chord progressions for 10 keyword sets under three conditions: 
(i) our rejection sampling approach (\system{}), (ii) keyword-conditioned LLM (GPT-4o), and (iii) LSTM trained on music data without keyword conditioning (LSTM Prior). 
For \system{}, we generated 30 4-bar chord progressions per keyword set using GPT-4o, applied rejection sampling, and randomly selected one from the accepted samples. For GPT-4o, we generated 30 4-bar chord progressions per keyword set and randomly selected one. For LSTM Prior, we generated a 4-bar chord progression using the LSTM model trained on HookTheory data~\cite{sheetsage} ($P(\mathbf{x})$). 
Each keyword set was processed five times, resulting in 10 keyword sets $\times$ 5 questions $\times$ \({}_{3}\text{C}_{2}\) pairs = 150 comparisons. Each task was evaluated by three unique listeners, yielding a total of 450 comparisons for each task (N=900 in total). 
\rev{Since the Shapiro-Wilk test indicated that the data was non-parametric, we analyzed these pairwise judgments using the Wilcoxon signed-rank test~\citep{conover1999practical}.}

We recruited 45 unique listeners from Prolific~\cite{prolific}, targeting individuals with a primary interest in music and 5+ years of experience in musical instruments. 
\rev{Participants were required to have audio devices and confirm they were in a quiet setting before starting the survey.}
Each participant answered 22 questions (10 on musical coherence, 10 on keyword relevance, and 2 attention checks) and was compensated \pounds 2, which corresponds to an approximate hourly rate of 12.13 USD.\footnote{Among the 49 participants initially recruited, four failed the attention checks and were excluded from the analysis.}

\paragraph{\textbf{Results.}}

\begin{figure*}[t!]
    \centering
    \begin{subfigure}[t]{0.45\linewidth}
        \centering
        \includegraphics[width=\linewidth]{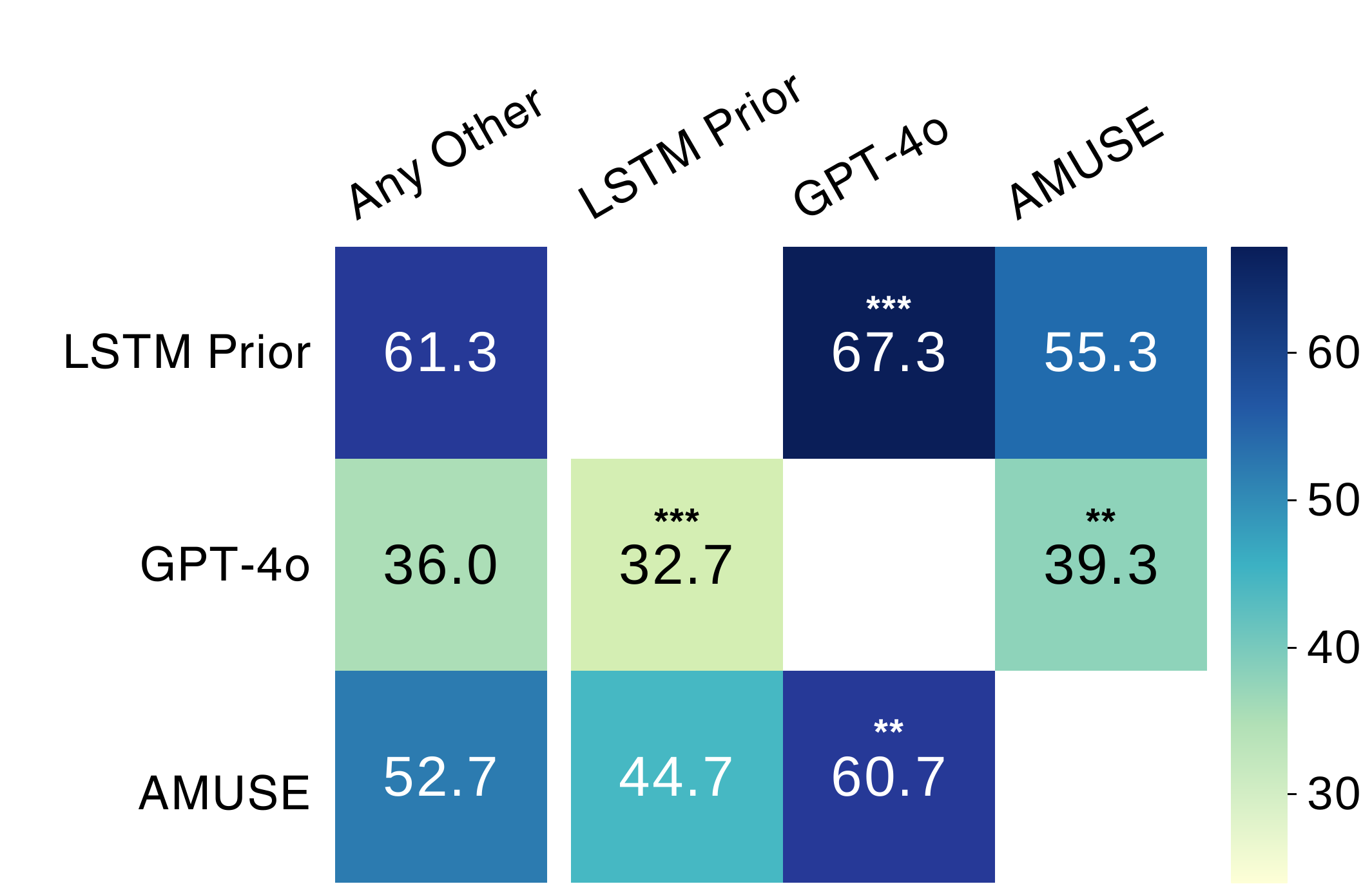}
        \caption{Musical Coherence}
        \label{fig:listener:coherence}
    \end{subfigure}
    \hfill
    \begin{subfigure}[t]{0.45\linewidth}
        \centering
        \includegraphics[width=\linewidth]{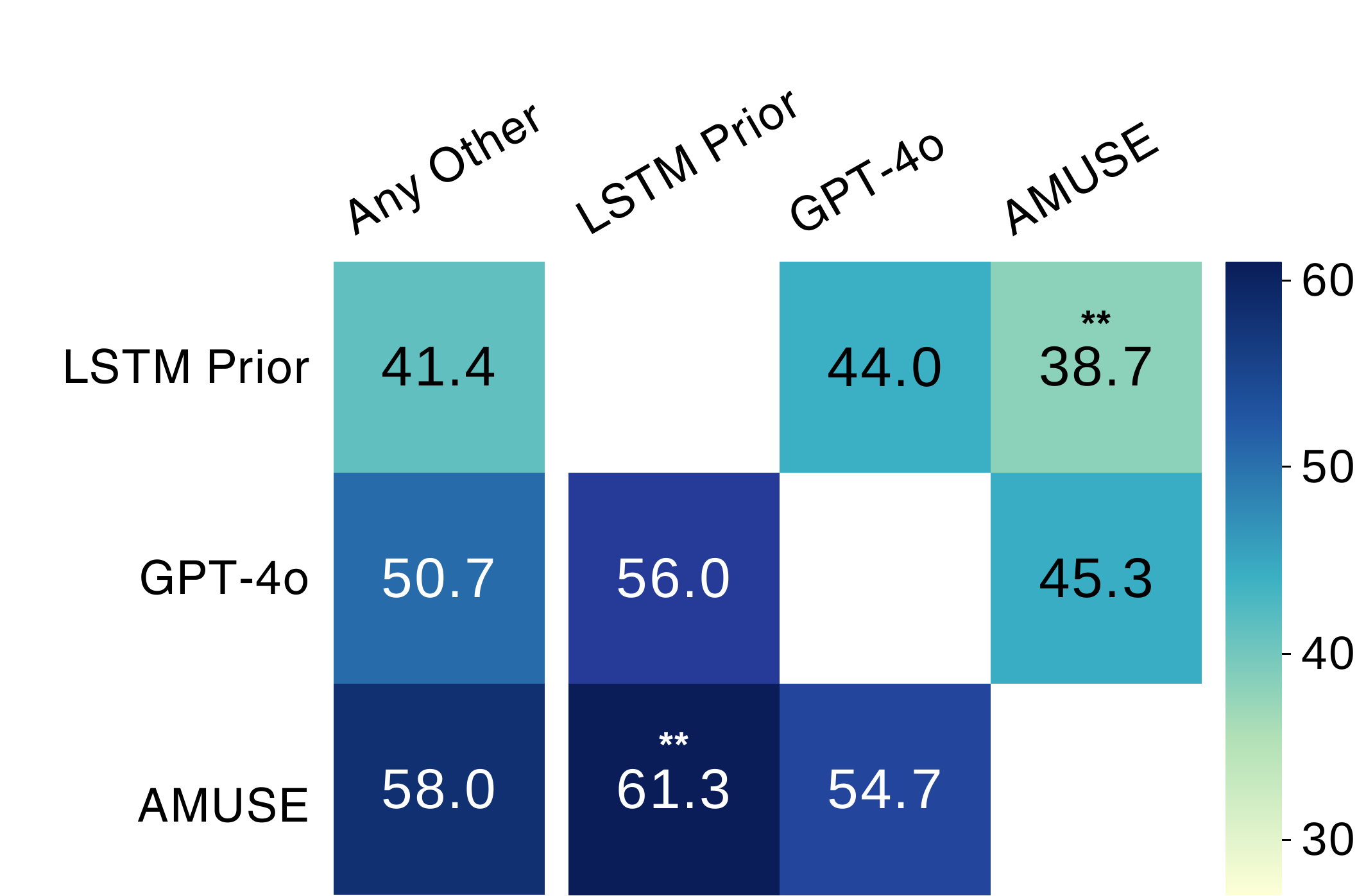}
        \caption{Keyword Relevance}
        \label{fig:listener:relevance}
    \end{subfigure}
    \caption{Results from our listening study where listeners indicated a preference between pairs of chord progression audio clips generated by two different methods among \texttt{LSTM Prior}, \texttt{GPT-4o}, or \texttt{\system{}}. Each row indicates the \% of times listeners preferred audio from that system compared to those from any other system (first column, N=300) and each system individually (other columns, N=150). 
    \rev{Wilcoxon signed-rank test with these paired data reveals that for Musical Coherence, \system{} shows no significant difference compared to LSTM Prior, which aligns closely with real music distributions (55.3\% of LSTM Prior generations preferred over \system{}). For Keyword Relevance, \system{} is most preferred by users (58\% of cases against any other samples), significantly outperforming LSTM Prior.} (**$p$<.01, ***$p$<.001).
    }
    \Description{
    This figure presents the results of a listening study that compares listener preferences between chord progression audio clips generated by three systems: LSTM Prior, GPT-4o, and AMUSE. For the musical coherence, the percentages show how often listeners preferred the system's output compared to any other system and to each individual system. AMUSE is preferred over GPT-4o and LSTM Prior in terms of musical coherence, receiving 60.7\% preference compared to GPT-4o (39.3\%) and LSTM Prior (44.7\%). For the keyword relevance, AMUSE is also generally preferred for keyword relevance, with a 61.3\% preference compared to LSTM Prior (44.0\%) and GPT-4o (45.3\%). Significant results are indicated, with p < 0.01 and *p < 0.001, suggesting strong statistical support for the findings in favor of AMUSE.
    }
    % \vspace{-1em}
    \label{fig:listener}
\end{figure*}

% We ensured the validity of our results by applying the Wilcoxon signed-rank test, a non-parametric method designed for paired samples, making it well-suited for small datasets [1].

Results for systems appear in Figure~\ref{fig:listener}. For Musical Coherence (Figure~\ref{fig:listener:coherence}), our method (\system{}) shows no statistical difference compared to LSTM Prior, which is the closest to real music distribution. 
Conversely, the analysis reveals that users did not prefer samples generated by GPT-4o in terms of musical coherence, with both LSTM Prior and \system{} being significantly more preferred than GPT-4o (LSTM Prior > GPT-4o: $p$=2.18$e$-5; \system{} > GPT-4o: $p$=0.009). 
For Keyword Relevance (Figure~\ref{fig:listener:relevance}), our system (\system{}) was the most preferred by users (58\% of cases against any other samples), significantly outperforming LSTM Prior ($p$=0.006). 

This result, combined with our quantitative findings on diversity and coherence, suggests that \system{} successfully achieves the desiderata we defined; our proposed method generates \textbf{diverse} and \textbf{coherent} chords that are \textbf{relevant} to the user input keywords.
\section{User Study}
\label{section:study}
We conducted a user study to gain insights and feedback on the potential, limitations, and future opportunities of multimodal music generation tools to support songwriting creativity. This overarching goal broke down into three main research questions:

\begin{enumerate}
    \item [\textbf{RQ1:}] How does \system{} support songwriters transforming multimodal inspirations into contextual musical elements?
    \item [\textbf{RQ2:}] How do songwriters incorporate \system{} into their creative workflows? 
    \item [\textbf{RQ3:}] How does \system{} impact the songwriting experience and perceived quality of the compositions?
\end{enumerate}

\subsection{Participants}
\label{section:study:participants}

We recruited 10 songwriters through targeted email outreach to active Hooktheory users, ensuring that our study included genuine songwriters actively engaging with the platform. We collaborated directly with Hooktheory to reach out to the users.
Eight identified themselves as hobbyists (i.e., write songs primarily for personal enjoyment) and two as professional (i.e., write songs as part of a career or generate income from songwriting). All were regular users of Hookpad and engaged in songwriting activity at least once a week. Six participants were regular users of \aria{} and four had not used \aria{} before. 
Participants composed in several different genres, including pop, rock, and blues. We detailed participants in Table~\ref{tab:main:participants}. Participants were compensated with 50 USD Amazon gift card for the 2-hour study.

\begingroup
\setlength{\tabcolsep}{8pt} % Default value: 6pt
\begin{table}[t]

\caption{Detailed background information of the participants in the user study (Section~\ref{section:study}). For songwriting proficiency, we use self-reported levels. Hobbyist = I write songs primarily for personal enjoyment; Professional = I write songs as part of my career or to generate income from my songwriting.}

\centering
\scalebox{0.9}{ % adjust the scale factor as needed
\begin{tabular}{c c l} %{@{}lcllll@{}}
\toprule
 \begin{tabular}[c]{@{}c@{}}Participant \\ ID\end{tabular}  & \begin{tabular}[c]{@{}c@{}}Songwriting \\ Proficiency\end{tabular} &  Songwriting Genre \\ 
 \midrule
P1 & Hobbyist & Blues, Rock, Americana \\
P2 & Hobbyist & Acoustic Pop, Symphonic \\
P3 & Hobbyist & New Age, Classical \\
P4 & Hobbyist & Indie, Alternative, Pop \\
P5 & Hobbyist & Rock, Pop, Country \\
P6 & Hobbyist & Pop \\
P7 & Hobbyist & Pop, Jazz Pop  \\
P8 & Hobbyist & Rock, Pop \\
P9 & Professional & Pop, R\&B, Hip Hop \\
P10 & Professional & Rock \\ 
\bottomrule
\end{tabular}
}

\label{tab:main:participants}
\vspace{-.5em}
\end{table}
\endgroup

\begin{figure*}[t!]
    \centering
    \includegraphics[width=\linewidth]{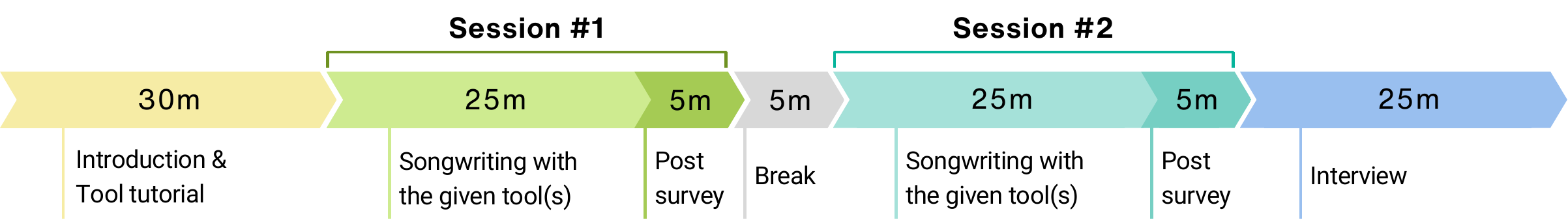}
    \caption{Overview of the user study procedure. 
    The study involved two songwriting sessions, each using a different set of tools. In each session, participants wrote an 8-bar chorus based on a given songwriting prompt. Following the sessions, participants took part in 25-minute semi-structured interviews about their experiences.
    }
    \Description{
    This figure presents an overview of the user study procedure. The study begins with a 30-minute introduction and tool tutorial. In Session #1, participants spend 25 minutes on songwriting using a specific tool, followed by a 5-minute post-survey. After a 5-minute break, Session #2 begins, where participants engage in another 25-minute songwriting session with a different tool, followed by another 5-minute post-survey. The study concludes with a 25-minute semi-structured interview to discuss the participants' experiences.
    }
    \label{fig:main:procedure}
\end{figure*}

\subsection{Study Procedure}
\label{section:study:procedure}

To connect with real-world Hooktheory users across the globe, we conducted a remote study using Zoom.
During the study, participants performed two songwriting tasks under different conditions: \baseline{} and \control{}. In the \baseline{} condition, participants were allowed to use only \aria{} (reflects the \textit{existing} AI-assisted songwriting practice), whereas in the \control{} condition, participants had access to both \system{} and \aria{} (introduces \textit{multimodal} inspiration support in addition to the conventional AI-assisted songwriting workflow). 
We opted not to compare to the manual composition without \aria{} because our primary goal was to evaluate the added benefit of \system{} in enhancing existing AI-assisted workflows rather than assessing AI versus non-AI approaches.
The task for each session was to write an 8-bar chorus (chords and melodies) given one of two songwriting prompts: ``Write an 8-bar chorus about the beginning of an unexpected friendship you had in your life'' or ``Write an 8-bar chorus about your favorite summer holiday memory.'' We adapted these prompts from the Songwriting subreddit~\cite{subreddit}. 
Both task prompts and system conditions were randomized across the sessions to minimize the ordering bias.

The overall study procedure is illustrated in Figure~\ref{fig:main:procedure}. 
We started with an introduction and tool tutorial session, where users were given 20 minutes to familiarize themselves with \system{} and \aria{}.
For each songwriting session, participants were given the songwriting prompt and had 25 minutes to write an 8-bar chorus using the assigned tool(s) with their screen shared. 
We asked participants to think aloud to learn their decision-making processes and reactions when interacting with the tools. The whole session, including the shared screen, was video recorded.
Since \system{} involves image inputs, we asked participants not to share any personal or sensitive pictures during the study.
To ensure meaningful comparison between conditions, we required participants to use the assigned tool(s) at least once during the task; however, they had the flexibility to decide when and how to use the features, as well as how frequently to engage with them.

\subsection{Measures}
\label{section:study:measures}

\subsubsection{Questionnaires}
\label{section:study:measures:questionnaires}
We analyzed participants' responses to the two post-task surveys. These surveys asked participants to rate, on a seven-point Likert scale, the usefulness of the given systems (\baseline{}: `\aria{} alone'; \control{}: `both \system{} and \aria{}') in helping transform their inspirations into music and writing songs aligned with the given song prompts. Additionally, participants were asked to assess their self-perceived experience with the system using questions adapted from related works~\cite{aichains, expressive}. 
We also collected participants' ratings on five aspects of the Creativity Support Index~(CSI)~\cite{csi} questionnaire, excluding the ``Collaboration'' question, as it is irrelevant to human-AI collaborations.
\rev{We analyzed these Likert scale ratings using the Wilcoxon signed-rank test, as a Shapiro-Wilk test indicated the data was non-parametric~\cite{conover1999practical}.} 
We include the detailed survey questions in Appendix~\ref{appendix:study:surveyquestions}.

\subsubsection{Interviews}
\label{section:study:measures:interviews}
During the 25-minute semi-structured interviews, 
participants were asked about the difference between the two songwriting sessions 
and the perceived impact of the tools on their songwriting tasks. 
Detailed interview questions can be found in Appendix~\ref{appendix:study:interviewquestions}. 

\rev{We transcribed the semi-structured interviews and analyzed them using the constant comparative method~\citep{strauss2017discovery}. 
Two authors independently open-coded two interview samples to identify key concepts and patterns. Axial coding was then performed to link these patterns~\citep{corbin2014basics}, resulting in an initial codebook.
The first author coded the remaining interview data while continuously refining the codebook.
After the first coding round, another author coded two interview samples using the updated codebook for verification, and any issues were resolved through discussion.
Throughout the process, the research team regularly discussed emerging themes while triangulating the interview data with quantitative analyses. We include the full codebook in Appendix~\ref{appendix:study:codebook}.
}
% Through open coding, we identifed a comprehensive taxonomy of
% linguistic strategies in gender debate, capturing fve overarching
% themes including derogation, gender distinction, intensifcation,
% mitigation, and cognizance guidance.

% The frst
% author open-coded all posts line-by-line. Then, the frst and third
% authors identifed frst-level themes from the open codes in weekly
% meetings over 2 months. Lastly, all authors met to refne the frstlevel themes and categorized them into second-level themes in
% weekly meetings over the course of 3 months

% For qualitative analysis, we followed the procedure of thematic analysis [9] and applied the
% constant comparative method [17]. We first transcribed the interviews conducted in Korean. Four
% authors individually coded two contrasting samples of the interviews and then discussed them
% together to share initial codes and potential themes, which were later turned into a codebook. The
% lead author coded the remaining interview data while continuously refining the codes and updating
% the codebook. Upon completion of the first round of coding, another author coded one interview
% sample based on the updated codebook for verification. Throughout this process, all authors met
% regularly to discuss potential themes while triangulating the interview data with the quantitative
% analyses. B

\subsubsection{Interaction Logs}
\label{section:study:measures:interactionlogs}
We collected the usage logs (i.e., participant actions with timestamps) to quantitatively analyze participant behaviors. We used this data to obtain statistics on tool usage, such as the time taken for each session, the number of keyword/chord generations in \system{}, and the number of chord/melody generations in \aria{}. For all these measures, we conducted a Shapiro-Wilk test to determine if the data was parametric and used a paired t-test for parametric data and a Wilcoxon signed-ranked test for nonparametric data. 
To qualitatively explain the user behaviors identified from the quantitative data, we analyzed screen recordings when necessary. 
\section{Findings}
\label{section:findings}

We present our findings on how participants transformed their multimodal inspirations into musical elements in \S\ref{section:findings:rq1} (RQ1), how \system{} influenced participants' songwriting processes in \S\ref{section:findings:rq2} (RQ2), and how \system{} impacted the songwriting experience and perceived quality of outcomes in \S\ref{section:findings:rq3} (RQ3). We summarize our main findings below:

\begin{itemize}
    \item [\textbf{RQ1:}] \system{} effectively supports the transformation of multimodal inspirations into concrete musical elements, with participants feeling more guided and aligned with their creative goals compared to using \aria{} alone.
    \item [\textbf{RQ2:}] Participants demonstrated diverse usage patterns of \system{} in the songwriting process, integrating it in various stages of their songwriting process. 
    \item [\textbf{RQ3:}] With \system{}, participants experienced enhanced agency and creativity throughout the songwriting process and perceived it as more efficient and easier. However, they found the compositions in both the \baseline{} and \control{} conditions to be equally satisfactory.
\end{itemize}

\subsection{Supporting Multimodal Inspirations}
\label{section:findings:rq1}

We first examine whether \system{} successfully supports songwriters transforming multimodal inspirations into contextual musical elements (RQ1). 
Specifically, we present the effectiveness of \system{} in the inspiration transformation process (\S\ref{section:findings:rq1:1}) and the perceived quality of suggestions (\S\ref{section:findings:rq1:2}). We also report the participants' preferences for the Chord Generator over Chord Transcriber within \system{} (\S\ref{section:findings:rq1:3}) and the impact of music keywords in the system usage (\S\ref{section:findings:rq1:4}).
We describe relevant quantitative findings with qualitative insights.

\subsubsection{\textbf{Effective transformation of multimodal inspirations into musical elements.}}
\label{section:findings:rq1:1}
Overall, participants in \control{} condition felt they could more easily transform their initial inspirations into concrete musical elements than \baseline{} (Inspiration Support in Figure~\ref{fig:findings:rating1}; \control=6.20$\pm$0.75, \baseline=4.60$\pm$1.36, $z$=0.00, $p$<0.01).
Additionally, participants reported feeling more guided toward the task goal in \control{} (Task Alignment in Figure~\ref{fig:findings:rating1}; \control{}=6.10$\pm$0.70, \baseline{}=4.30$\pm$1.62, $z$=0.00, $p$=0.016).
Six participants emphasized how \textbf{\system{} helped better achieve their creative goals}. \system{}'s suggestions ``\textit{fell in line with what I would have mentally constructed}'' (P5), allowing them to ``\textit{focus more on the goal with a clear vision of how to write the rest of the song}'' (P2). % (P1-2, P5-8) 
In contrast, in \baseline{} condition, five participants mentioned feeling unsupported in achieving the task, highlighting a lack of direction. P8 stated, ``\textit{Using \aria{} alone, I had a very vague vision and didn't really know what I was going for.}'' Similarly, P1 noted that without \system{}, ``\textit{the song ended up being the opposite of the style I originally intended to create with my lyrics}''.
% (P1, P2, P5, P7-8)

\subsubsection{\textbf{Higher perceived quality of suggestions from \control{}}}
\label{section:findings:rq1:2}
\begin{figure}[t!]
    \centering
    \includegraphics[width=\linewidth]{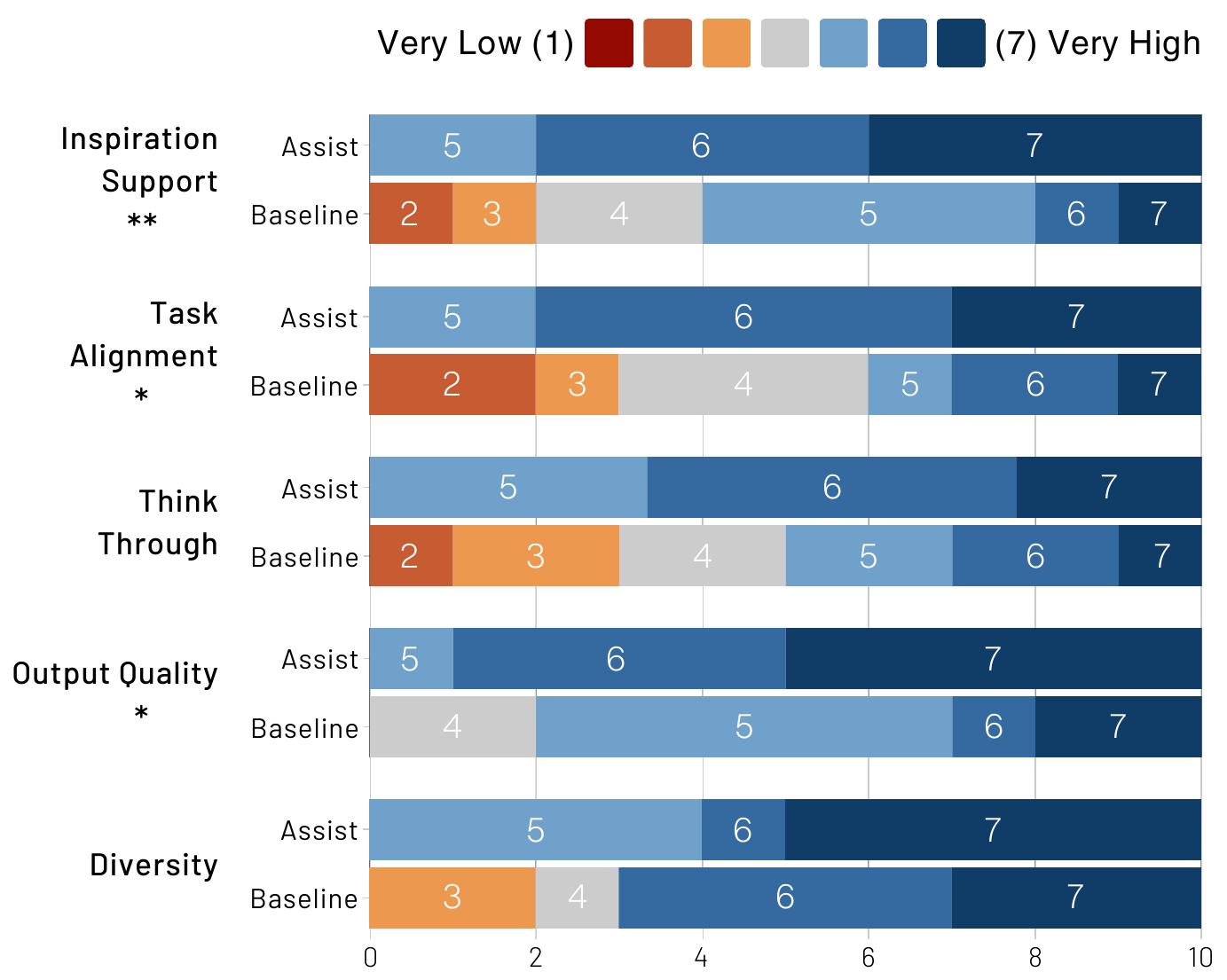}
    \caption{Distribution of participants' ratings on self-perceived songwriting experience. 
    When equipped with \system{}, participants felt they could more easily translate multimodal inspirations into musical elements (Inspiration Support), align their compositions with the given tasks (Task Alignment), and produce output of comparable quality (Output Quality). (*$p$<.05, **$p$<.01)
    }
    \Description{
     This figure shows the distribution of participants' self-reported ratings for their songwriting experience when using AMUSE (Control) compared to the Baseline. The ratings cover five dimensions: Inspiration Support, Task Alignment, Think Through, Output Quality, and Diversity, with scores ranging from 1 (Very Low) to 7 (Very High). Participants rated AMUSE significantly higher for Inspiration Support and Task Alignment compared to the Baseline. Output Quality was also rated higher for AMUSE, indicating participants felt it helped them produce comparable quality outputs. Ratings for Think Through and Diversity showed no significant difference between the two conditions.
    }
    \label{fig:findings:rating1}
    \vspace{-1em}
\end{figure}

While participants expressed satisfaction with the diversity of suggestions in both conditions (Diversity in Figure~\ref{fig:findings:rating1}; \control=6.10$\pm$0.94, \baseline=5.50$\pm$1.50; $z$=12.00, $p$=0.389), participants \textbf{rated the quality of the outputs they received from \control{} significantly higher} than those they got from \baseline{} (Output Quality in Figure~\ref{fig:findings:rating1}; \control=6.40$\pm$0.66, \baseline=5.30$\pm$1.00, $z$=4.00, $p$=0.023). 
Interestingly, there was no significant difference in the average number of \aria{} queries made between the conditions (\baseline=3.60$\pm$2.72, \control=2.30$\pm$1.16; z=10.0, p=0.130) or in the acceptance rates of those queries (\baseline=0.71$\pm$0.25, \control=0.87$\pm$0.22, z=7.0, p=0.12). 
This similarity suggests that the improved perceptions of output quality may be more related to \system{}'s ability to generate suggestions that align with user intentions, leading to increased general satisfaction with the suggestions received throughout the songwriting process.

\subsubsection{\textbf{Preference for Chord Generator over Chord Transcriber}}
\label{section:findings:rq1:3}
Participants demonstrated a clear preference for the Chord Generator over the Chord Transcriber within \system{}. 
All participants utilized the Chord Generator, whereas 
only three used the Chord Transcriber in \control{} ($p$ < 0.001).
Five participants praised the Chord Generator, \textbf{describing the chords created with multimodal inspirations as ``\textit{special}''} % (P4, P6) 
compared to those generated in \baseline{}. % P2, 4, 5, 6, 8
Three participants found the \textbf{Chord Transcriber less creative than Chord Generator}, with P5 noting, ``\textit{With Chord Transcriber, you are picking pieces someone else created---it is not mine and feels close-ended. There’s a risk of sounding like something that already exists. But the Chord Generator feels more open-ended and freeing.}''
P6 suggested making Chord Transcriber's output unique by integrating it with the Chord Generator: ``\textit{Transcribing the chords from the song I love would be good, but what I find challenging is changing these transcribed chords to make them your own. It would be cool if you could input the Chord Transcriber's outputs into the Chord Generator to create new chords that match the vibe of the original song.}''

The low usage of Chord Transcriber might also be attributed to the task constraints, as two participants noted. P2 mentioned they could not think of a song to transcribe given the songwriting prompt: ``\textit{I didn't use the transcribe feature which I think mainly because of the time constraint for me---I couldn't come up with a song that kind of fit the task that I could borrow from off the top of my head.}'' 
P10 did not use the Chord Transcriber due to the absence of their instrument, stating, ``\textit{I would have used it if I had my piano next to me. I would come up with whatever I felt and then upload the sound file to get the chords.}''

\subsubsection{\textbf{Enhanced transparency, agency, and explainability through music keywords}}
\label{section:findings:rq1:4}
Five participants found the keywords helpful in the transformation process, particularly for the \textbf{transparency and agency the keywords offered}. Participants appreciated how the keyword control allowed them to see what the AI was interpreting from text or images, select keywords in consensus, and regenerate or edit them to better align with their creative goals.
P5 noted, ``\textit{Taking an image or story and turning it directly into chord progressions is an abstract jump, but with the keywords, the process becomes intuitive and predictable.}'' P2 added, ``\textit{Keywords put into words the idea that I was going for.}''
\textbf{Keywords also provided explainability for chord outputs}. P6 explained, ``\textit{When the chord progressions were weird, I realized I was using the keyword `unexpected.' When I removed it, the generated chords were great.}''
These findings reconfirm that the keywords could be an effective communication medium for connecting abstract relationships between different modalities as suggested in prior work~\citep{opal, 3dalle}.

\begin{figure*}[t!]
    \centering
    \includegraphics[width=.78\linewidth]{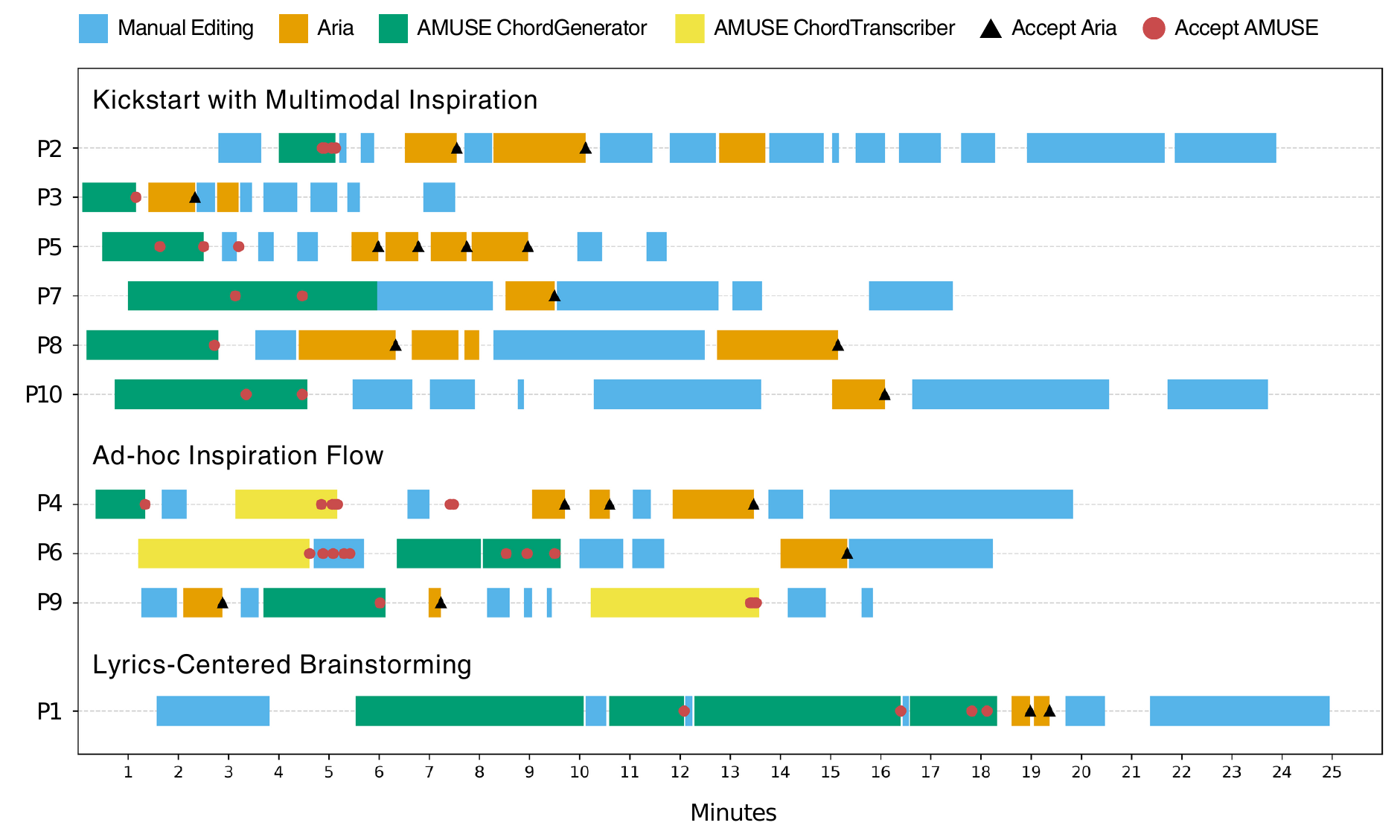}
    \vspace{-.5em}
    \caption{
    Interaction log timelines of all participants in the \control{} condition. Interaction patterns were categorized into three distinct approaches: (i) kickstart with multimodal inspiration, (ii) ad hoc inspiration flow, and (iii) lyrics-centered brainstorming. `Manual Editing' refers to user-initiated edits inside the Hookpad editor, including adjustments to melody, chords, instrumentation, and lyrics. `Aria' refers to instances where participants queried \aria{} to generate melodies or chords. `\system{} ChordGenerator' and `\system{} ChordTranscriber' refer to instances where participants were using each functionality. `Accept Aria' and `Accept \system{}' refer to events where participants accepted each tool's suggestions, respectively. Empty spaces between the color bars indicate periods when the user was either listening to the composed song without taking any action, or explaining their thoughts out loud as part of the think-aloud protocol, again without taking any direct actions.
    }
    \Description{
    This figure presents a timeline visualization of participants' interactions with different tools during the songwriting process, categorized into three approaches: Kickstart with Multimodal Inspiration, Ad-hoc Inspiration Flow, and Lyrics-Centered Brainstorming.
    }
    \vspace{-.5em}
    \label{fig:findings:controltimeline}
\end{figure*}

\subsubsection{\textbf{Summary of findings}}
% \subsubsection{\textbf{Answer to RQ1: How does \system{} support songwriters transforming multimodal inspirations into contextual musical elements?}}
\label{section:findings:rq1:summary}
\rev{The findings demonstrate that \system{} effectively bridges the gap between multimodal inspirations and musical elements.
Compared with using \aria{} alone, participants felt more guided and found it easier to achieve their creative goals with \system{}. 
The Chord Generator emerged as a preferred tool, offering unique chord progressions that participants found more inspiring than those from the Chord Transcriber. 
Positive user comments indicate that the music keyword pipeline further enhanced the experience by providing transparency, explainability, and control over the AI's output. 
Collectively, these findings confirm that 
\system{} 
effectively supports songwriters in 
transforming their
multimodal inspirations into contextual musical elements.}

\subsection{Impact on User's Songwriting Process}
\label{section:findings:rq2}

We describe how participants incorporated \system{} into their songwriting process (RQ2). 
We first describe three \system{} usage patterns observed in \control{} condition (\S\ref{section:findings:rq2:control}), and explain how participants completed the task without \system{} in \baseline{} condition, using only \aria{} (\S\ref{section:findings:rq2:baseline}).
We report qualitative findings from analyzing screen recordings, think-aloud statements, and interviews. 

\subsubsection{\textbf{Diverse usage patterns of \system{} in the songwriting process.}}
\label{section:findings:rq2:control}

In the \control{} condition, we identified three categories of approaches that participants used to find and integrate inspiration into their songwriting process: (i)~kickstart with multimodal inspiration (N=6; P2-3, P5, P7-8, P10), (ii)~ad hoc inspiration flow (N=3; P4, P6, P9), and (iii)~lyrics-centered brainstorming (N=1; P1). 
Participants engaged with \system{} at diverse stages of their workflow, with its use generally concentrated in the early stages.
In addition, when using \system{}, participants reported a shift in their approach to \aria{}, using it more for melody creation rather than for both chords and melody.
The timeline of all participants' songwriting patterns is shown in Figure~\ref{fig:findings:controltimeline}. 

\begin{figure*}[t!]
    \centering
    \includegraphics[width=.8\linewidth]{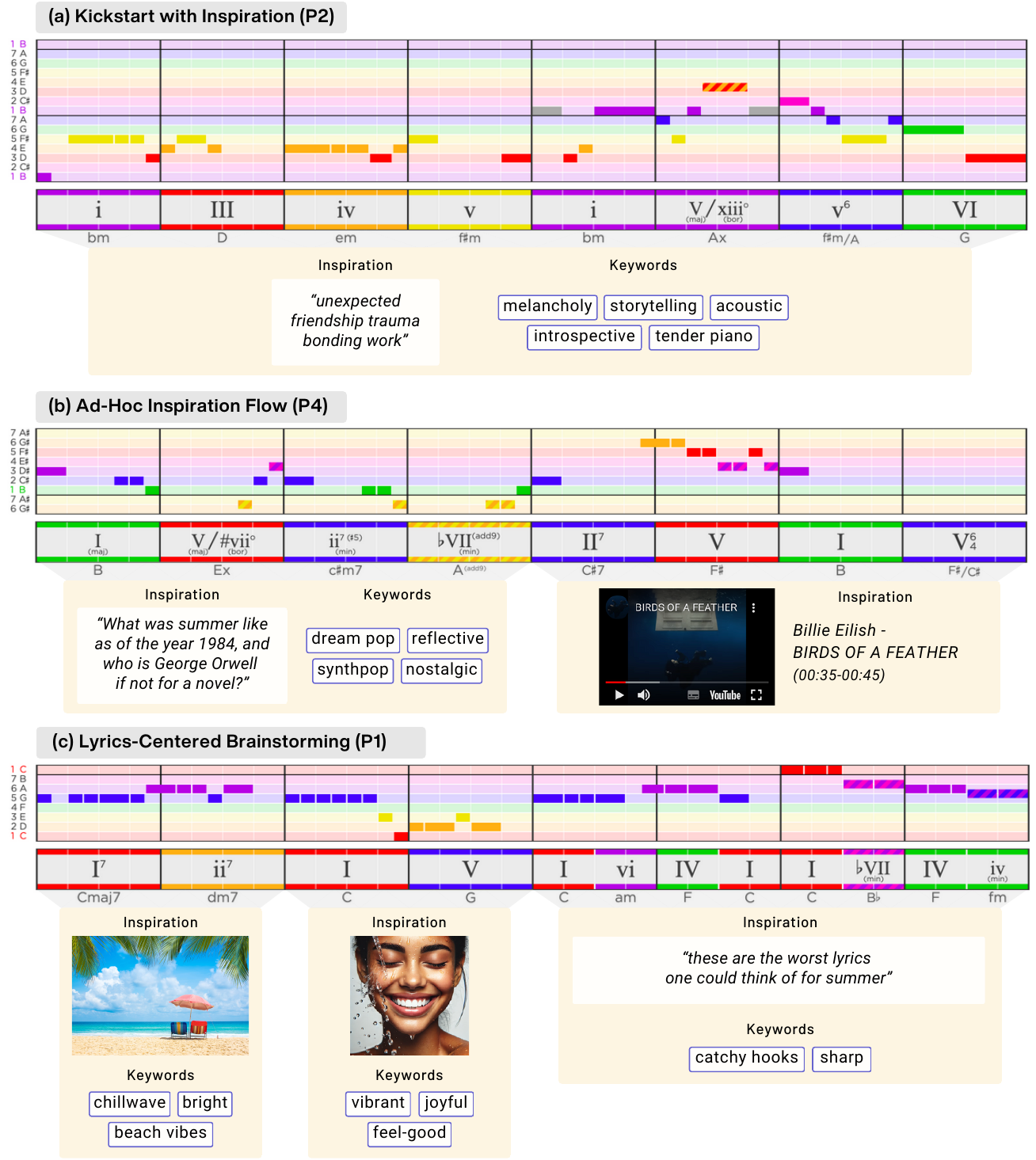}
    \caption{
    Three different songwriting approaches in the \control{} condition. (a) Kickstart with Inspiration: P2 generated chord progressions with \system{} in the early stages of the process, creating the full chord progression before moving forward with other elements. (b) Ad hoc Inspiration Flow: P4 first used \system{} with a text-based inspiration, which reminded them of Billie Eilish's song Birds of a Feather. P4 returned to \system{} to build upon that inspiration. (c) Lyrics-Centered Brainstorming: P1 mapped out the lyrical content for each bar and used \system{} to generate corresponding musical elements for each section.
    }
    \Description{
    This figure illustrates three different songwriting approaches using AMUSE in the Control condition: (a) Kickstart with Inspiration (P2): P2 quickly generated chord progressions using AMUSE at the early stages of the song, creating a full progression (e.g., bm - D - em - f#m) and then moved on to add other elements. The keywords selected included "melancholy," "storytelling," and "acoustic." (b) Ad-hoc Inspiration Flow (P4): P4 started with AMUSE based on a text-based inspiration, which reminded them of a segment from Billie Eilish's song Birds of a Feather (used for further inspiration). Keywords like "dream pop," "reflective," and "synthpop" guided the chord progression creation. (c) Lyrics-Centered Brainstorming (P1): P1 mapped lyrical content for each bar and used AMUSE to generate matching musical elements for each section. Keywords such as "vibrant," "joyful," and "feel-good" were used to inspire the chord progression.
    }
    \label{fig:findings:amuseillustration}
\end{figure*}

\paragraph{\textbf{Kickstart with Multimodal Inspiration.}}
Given the songwriting prompt, six participants (P2-3, P5, P7-8, P10) \textbf{quickly drew on personal inspirations and used \system{}} to generate chord progressions that matched their ideas. After the initial generation with \system{}, they did not return to \system{}; instead, they developed the rest of the song by expanding or editing the chord progressions created with \system{} and composing melodies, either manually or using \aria{}. 
For example, P2 was inspired by a recent `friendship at work created through a trauma bonding experience' and aimed to write a melancholy, storytelling song (see Figure~\ref{fig:findings:amuseillustration}(a) for the illustration). P2 accepted the four 4-bar chord progressions generated by \system{} and combined them to create the entire chord progression of the song. They then developed the rest of the song by generating melodies with \aria{} and editing those.

\paragraph{\textbf{Ad hoc Inspiration Flow}}
In contrast to users who used \system{} only at the first stage to kickstart their songwriting process, three participants (P4, P6, P9) constantly used \system{} in their workflow. 
P4 and P6 mentioned that \textbf{transforming their initial inspiration into chords with \system{} sparked further ideas.}
For instance, P4 described starting with inspiration from George Orwell's \textit{1984} and then transitioning into Billie Eilish's \textit{Birds of a Feather} (see Figure~\ref{fig:findings:amuseillustration}(b) for the illustration): ``\textit{1984 is a book about conformity and structure, and I hoped to capture its serenity and finer ambiance. I thought keywords like `dream pop' could aesthetically intertwine with my inspiration. As I listened to the chords, I wanted the next four bars to progress into a darker, non-stereotypical mood. Billie Eilish's Birds of a Feather came to mind because of its down-tempo, dreamy ambient soundscape, which would create a good resolution in bars five through eight.}''

\begin{figure*}[t!]
    \centering
    \includegraphics[width=.78\linewidth]{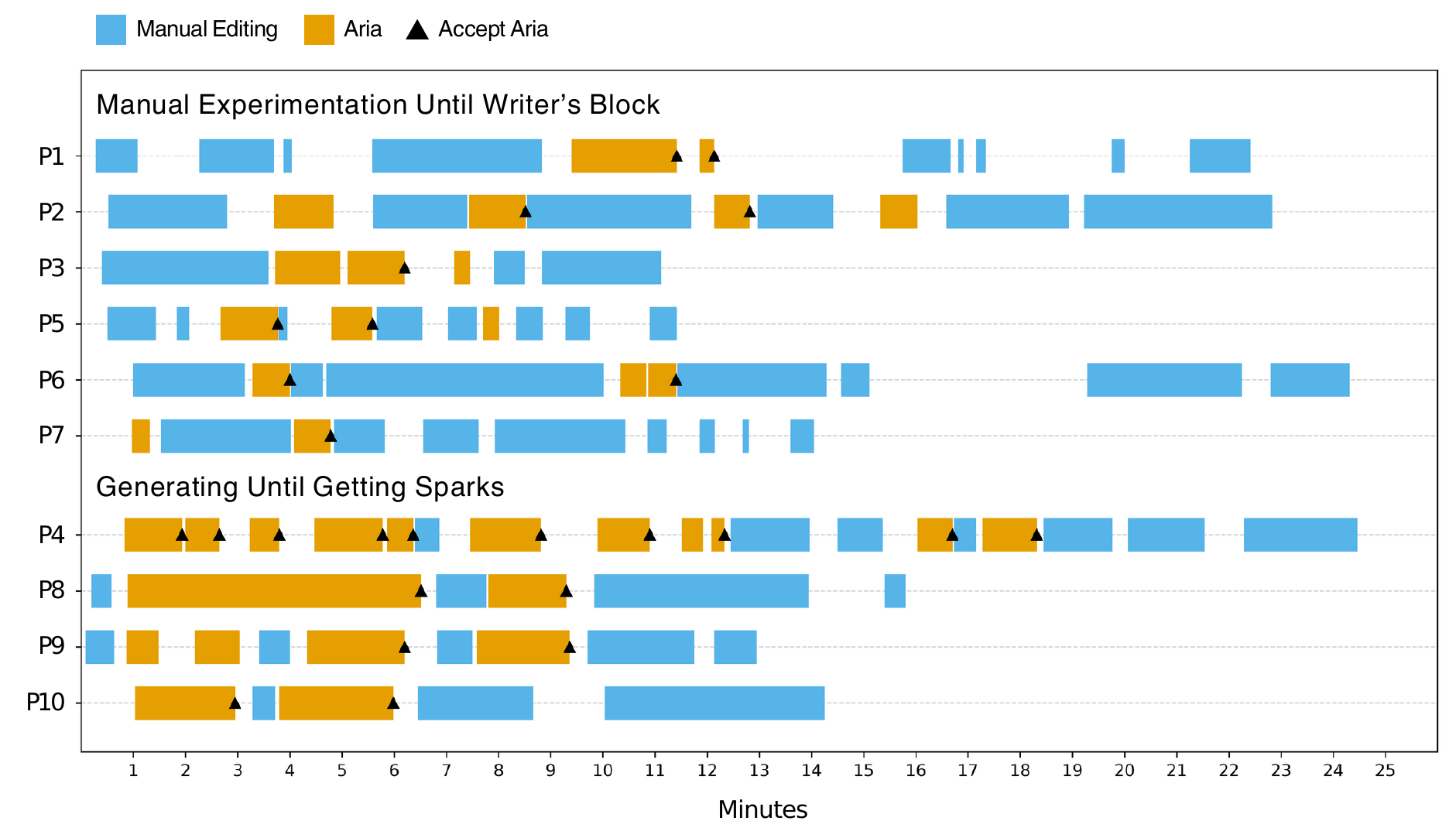}
    \vspace{-.5em}
    \caption{
    Interaction log timelines of all participants in the \baseline{} condition. Interaction patterns were categorized into two distinct approaches: manual experimentation until writer’s block and generating until getting sparks. `Manual Editing' refers to user-initiated edits inside the Hookpad editor, including adjustments to melody, chords, and lyrics. `\aria{}' refers to instances where participants queried \aria{} to generate melodies or chords, and `Accept \aria{}' refers to events were participants accepted \aria{}'s suggestions. Empty spaces between the color bars indicate periods when participants were either listening to the composition without making changes or reflecting on their creative process as part of the think-aloud protocol, without engaging in direct editing.
    }
    \Description{
    This figure shows two different approaches participants used when working with Aria during the songwriting task: Manual Experimentation Until Writer’s Block and Generating Until Getting Sparks.
    }
    \vspace{-.5em}
    \label{fig:findings:baselinetimeline}
\end{figure*}

P9, on the other hand, took a different approach and was not fully satisfied by \system{}'s generations. 
Rather than starting with \system{}, P9 manually composed the chord progression for bars 1-4 and completed the melody with \aria{}. 
The inspiration to reflect the mood of `enjoying a 4th of July fireworks' came \textbf{midway through the songwriting process}, leading P9 to use \system{}'s Chord Generator for the last four bars. Later, P9 drew inspiration from Childish Gambino's song \textit{Saturday} to infill the chord at bar 5, where he realized the chords for bars 1-4 and 5-8 did not fit well together. This led P9 to manually rewrite the progression for bars 5-8. 
Reflecting on the experience, P9 said, ``\textit{I could easily come up with the chord progression for the first few bars. I wanted to tell a story about the 4th of July fireworks using \system{} and expected it to build on the chords I had already laid down, but I didn't like the result.}''
Unlike \aria{}, \system{} does not have access to the full context of the user's composition, and we speculate that providing context would be necessary to better support users in the ad hoc inspiration flow.

\paragraph{\textbf{Lyrics-Centered Brainstorming.}}
One participant, P1, divided the song into three sections, \textbf{assigning lyrical ideas to each part and using \system{} to generate chords} that aligned with those ideas (see Figure~\ref{fig:findings:amuseillustration}(c) for the illustration). P1 began by outlining the lyrical ideas, noting: `Warm breeze salt in the air, cool water pouring over your hair, these are the worst lyrics one could think of for summer.'
They then structured the song by planning bars 1-2 around a `gentle summer beach,' bars 3-4 for `cool water pouring over your hair,' and bars 4-8 to sing about the `worst lyrics' theme. For the first two sections, P1 searched for relevant images online and used the Chord Generator to get chords matching the lyrical concepts. For the last four bars, he entered the lyric to the Chord Generator and selected two four-bar chord progressions that are singable. \aria{} was used later in the process to assist with melody creation as P1 worked to bring the composition together. 
While most users relied on visual inputs to \system{}, the instincts of this particular user centered strongly around lyrics, providing further evidence that systems should accommodate diverse multimodal inputs (DG1) to support a variety of users.

\subsubsection{\textbf{Songwriting approaches without \system{}}}
\label{section:findings:rq2:baseline}

In \baseline{} condition, where the participants wrote a song without \system{}, participants either (i) manually experimented with chords and melodies until encountering writer's block (N=6; P1-3, P5-7), or (ii) relied on \aria{} from scratch to find generations that align with inspiration (N=4; P4, P8, P9, P10). 
The timeline of all participants' songwriting process is shown in Figure~\ref{fig:findings:baselinetimeline}.

\paragraph{\textbf{Manual Experimentation Until Writer's Block.}} 
Six participants (P1, P2, P3, P5, P6, P7) began by creating chords or melodies manually. This process involved \textbf{experimenting with different combinations of musical elements} to meet their creative goals. Once they had set up the desired mood, they turned to \aria{} for additional suggestions when they felt stuck. 
For example, P3 manually set up the full chord progressions and composed the melodies for the first five bars. Hoping ``\textit{\aria{} will finalize the song,}'' P3 used it to generate the melodies for the last three bars. 
Different from the others (P1-3, P5-6) who started writing manually from the beginning, P7 first tried using \aria{} for chords but found ``\textit{the chords do not sound summery.}'' Frustrated, P7 canceled generation and set up the full chord progressions manually and used \aria{} later for melody generation.

\paragraph{\textbf{Generating Until Getting Sparks.}}
Four participants (P4, P8, P9, P10) relied on \aria{} to spark their initial inspiration before writing any music manually, hoping to get a suggestion that aligned with their initial inspirations. 
However, this process often \textbf{resulted in suboptimal outputs that did not align with their intentions}.
For example, P10 planned to write about `the beginning of an unexpected friendship with my dog' but found that \aria{}'s chord suggestions were ``\textit{cool but bizarre for my dog.}'' After more attempts, P10 settled on a chord progression that ``\textit{didn't sound like a chorus, but was okay-ish,}'' and then manually adjusted it to better fit their goal.

\subsubsection{\textbf{Summary of findings}}
% \subsubsection{\textbf{Answer to RQ2: How do songwriters incorporate Amuse into their creative workflows?}}
\label{section:findings:rq2:summary}
\rev{
Based on the formative study findings, \system{} was designed to support the early stages of songwriting by helping users brainstorm and transform initial ideas into musical elements.
Interestingly, participants exhibited diverse usage patterns, leveraging \system{} flexibly across different stages of their songwriting process rather than exclusively at the beginning---such as during ad hoc inspiration flows or after brainstorming lyrics---with \aria{} often used to generate melodies on chords created with \system{}.
In contrast, without \system{}, users relied solely on \aria{} to generate both melodies and chords to ease cognitive inertia in the process, often receiving suggestions that did not align with their creative intentions. 
These findings demonstrate how multimodal control prompts users to 
integrate inspirations dynamically, and thus, 
the importance of supporting multimodalities \textit{throughout} the process 
to meet a wide range of creative workflows.}

\subsection{User Perception on Songwriting Process and Outcomes} 
\label{section:findings:rq3}

We investigate \system{}'s impact on the songwriting experience and perceived quality of the compositions (RQ3).
Concretely, we discuss \system{}'s impact on participants' sense of agency (\S\ref{section:findings:rq3:1}), creativity (\S\ref{section:findings:rq3:2}), and efficiency (\S\ref{section:findings:rq3:3}) in the songwriting process. Lastly, we compare the overall satisfaction with final compositions across both conditions (\S\ref{section:findings:rq3:4}). 
We describe relevant quantitative findings with qualitative insights.

\subsubsection{\textbf{Enhanced agency over the songwriting process}}
\label{section:findings:rq3:1}
\begin{figure*}[t!]
    \centering
    \includegraphics[width=\linewidth]{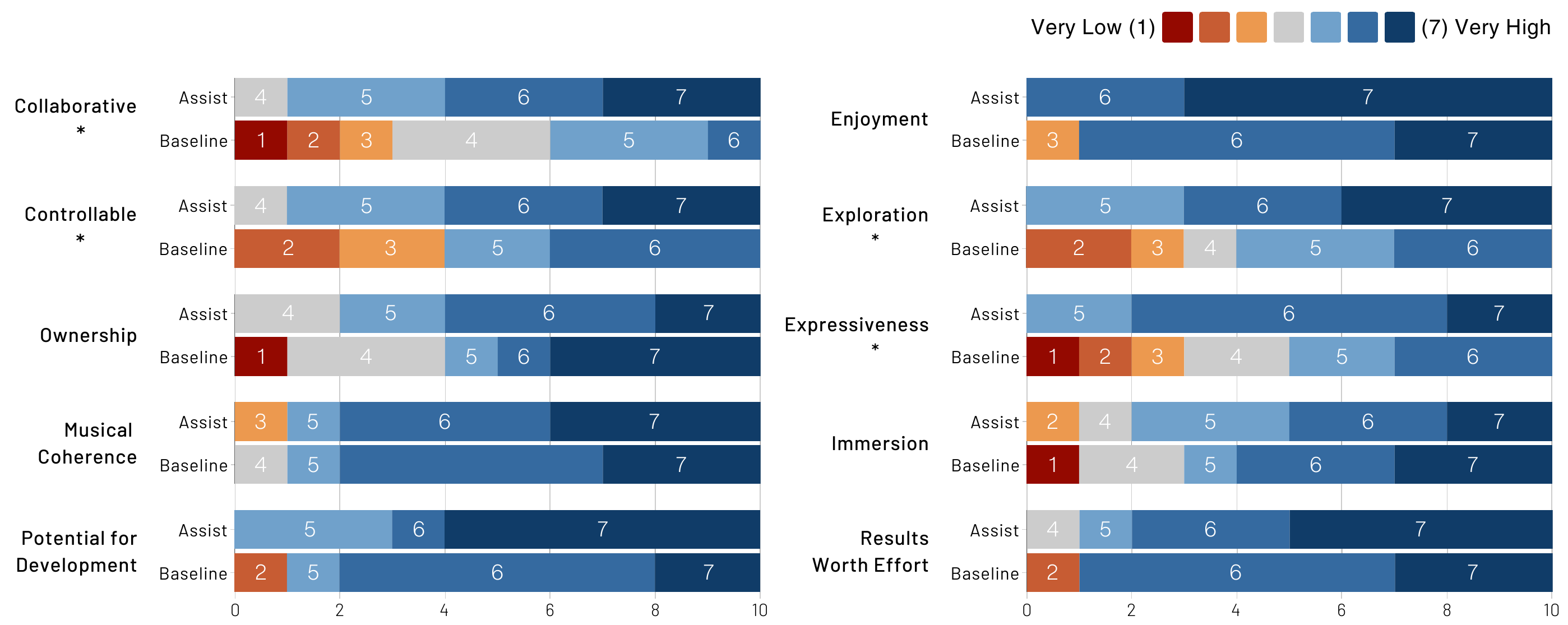}
    \caption{
    Distribution of participants' ratings on self-perceived songwriting experience and Creativity Support Index (CSI). The left side focuses on AI-related aspects and composition output, while the right side represents CSI questionnaires. 
    When equipped with \system{}, participants felt that they had more control in songwriting processes, and felt more collaborating with the tool. In addition, participants felt more expressive during the \control{} with more options to explore (*$p$<.05).}
    \Description{
    This figure shows participants' ratings on self-perceived songwriting experience and Creativity Support Index (CSI) across two conditions: Control (with AMUSE) and Baseline.The left side focuses on AI-related aspects and composition output: Collaborative and Controllable were rated significantly higher in the Control condition, indicating participants felt more collaboration and control when using AMUSE. Other dimensions such as Ownership, Musical Coherence, and Potential for Development show no significant differences but generally higher ratings in the Control condition. The right side represents the CSI aspects: Enjoyment, Exploration, and Expressiveness were also rated higher in the Control condition, showing that participants felt more expressive and exploratory with AMUSE, along with a sense of enjoyment.Immersion and Results Worth Effort showed minimal differences between the two conditions.
    }
    \label{fig:findings:rating2}
\end{figure*}

Overall, participants felt more in control over the process in \control{} compared to the \baseline{} (Controllable in Figure~\ref{fig:findings:rating2}; \control=5.80$\pm$0.98, \baseline=4.40$\pm$1.62, $z$=5.00, $p$=0.036). They also felt the process as more collaborative (Collaborative in Figure~\ref{fig:findings:rating2}; \control=5.80$\pm$0.98, \baseline=3.90$\pm$1.45, $z$=2.00, $p$=0.014), indicating that the tools in the  \control{} condition fostered a stronger sense of partnership during the songwriting process compared to the \baseline{} condition. 
Five participants mentioned that they could \textbf{communicate their creative intentions} and have \textbf{more agency over the workflow} when equipped with \system{}. %1, 2, 8, 10, 7
For instance, P1 said, ``\textit{I can be part of the creative process with \system{}. You can influence what it generates because you get to tell it what you want it to consider.}''
In contrast, two mentioned in \baseline{} they were simply ``\textit{following what \aria{} wants to do}'' (P2, P10). P10 explained, ``\textit{With just \aria{}, I wasn't able as much to create a mood or feel for my song. I didn't feel like I had much input---\aria{} just sort of made up its own mind, and I had to just go along with it. It's like having a bad co-writer.}''

An interesting finding from the interviews, mentioned by three participants, was that \textbf{having greater control during the early stages of songwriting with \system{} led to a sense of disconnection later in the process}. 
Participants expressed that \textbf{they set higher expectations for \aria{},} particularly regarding the relevance of its suggestions to their sources of inspiration after receiving suggestions from \system{}.
P3 mentioned, ``\textit{Now that the chords are about a boy playing in the garden, I wished \aria{} to generate something playful-maybe a melody with shorter notes and a quicker rhythm that would capture the image of a boy jumping. That was what I anticipated, but it didn't quite work.}''
Similarly, P4 noted that \textbf{using \system{} and \aria{} together felt less intuitive}: ``\textit{I felt like the chords were in place all the way through using \system{}. But the only disconnection I felt was in how \aria{} doesn't necessarily integrate with \system{} in the same way that \system{} integrates with \aria{}. So it was almost like doing a 180-degree rotation when switching from \system{} to \aria{} because \aria{} doesn't achieve quite the same results, given that it doesn't consider your inputs. [...] So I would say when \system{} was there, I felt more enriched and musical, but using \aria{} alone was more intuitive because I knew what to expect from it.}''
These findings suggest that both systems---\system{} and \aria{}---would benefit from incorporating each other's control inputs, i.e., contextual and multimodal inputs, respectively.

\subsubsection{\textbf{Enhanced creativity with multimodal inspirations}}
\label{section:findings:rq3:2}

When using both \system{} and \aria{} in the \control{} condition, participants 
reported a greater sense of idea exploration (Exploration in Figure~\ref{fig:findings:rating2}; \control=6.10$\pm$0.83, \baseline=4.40$\pm$1.50, $z$=3.00, $p$=0.019) as well as expressiveness and creativity (Expressiveness in Figure~\ref{fig:findings:rating2}; \control=6.00$\pm$0.63, \baseline=4.20$\pm$1.66, $z$=4.00, $p$=0.028) compared to the \baseline{}. These results indicate that \system{} enhanced participants' capacity to \textbf{explore new musical ideas} and \textbf{freely express their creative intentions}.
This finding echoed in the interviews, where seven participants praised \system{} for fostering creativity and offering greater freedom in musical expression.
%2, 6, 10, 4, 8, 1, 7 
P4 and P7 described \system{} as ``\textit{liberating}'' their creative process; P4 said, ``\textit{I feel like I'm actually guiding the tool toward what I want to create in a transformative, imaginative, and visually compelling world, almost as if I'm creating music as I type.}'' In addition, P2 appreciated the integration of visuals in the songwriting process, saying, ``\textit{I especially liked the picture input feature because, as the saying goes, a picture is worth a thousand words. Sometimes, it's easier to convey what you're aiming for with an image than with words.}''

\subsubsection{\textbf{Perceived efficiency in the songwriting process}}
\label{section:findings:rq3:3}
Participants reported that the songwriting process felt more efficient in \control{} during the interviews, despite no significant difference in actual completion times between conditions.
Seven participants noted that the process \textbf{felt more efficient and easier} in the \control{} condition.
They described how \system{} helped to ``\textit{streamline the process}'' (P8, P9), contrasting it with their usual experience where ``\textit{it takes a while before getting a chord progression that I love}'' (P6). P2 complimented, ``\textit{Honestly, I would use \system{} almost every time. I really liked how it jump-started the process. I usually spend a lot of time and effort getting the right chords, which is tedious—being able to convey an idea quickly was really nice.}''
However, the perceived efficiency \textbf{did not mean using \system{} really sped up the process}-there was no statistical difference in the task completion time between \control{} and \baseline{} (\control=17.39$\pm$5.45 min, \baseline= 17.61$\pm$5.27 min, $z$=26.00, $p$=0.922). This suggests that while \system{} may enhance the perceived ease of the songwriting process, particularly by alleviating specific pain points like finding the right chords, it does not necessarily make the overall task faster. 

\subsubsection{\textbf{Similarly satisfactory results in both conditions}}
\label{section:findings:rq3:4}
Despite the reported differences in agency and creativity in the songwriting process, participants in both conditions often arrived at satisfying results. 
The post-survey results revealed that participants rated the quality of their compositions similarly in both conditions: they felt that the results were worth the effort (Results Worth Effort in Figure~\ref{fig:findings:rating2}; \control=6.20$\pm$0.98, \baseline=5.90$\pm$1.37, $z$=9.00, $p$=0.750), musically coherent (Musical Coherence in Figure~\ref{fig:findings:rating2}; \control=6.00$\pm$1.18, \baseline=6.00$\pm$0.89, $z$=10.50, $p$=1.00), and could be expanded to full songs (Potential for Development in Figure~\ref{fig:findings:rating2}; \control=6.30$\pm$0.90, \baseline=5.70$\pm$1.35, $z$=12.50, $p$=0.429). Moreover, participants found both conditions enjoyable (Enjoyment in Figure~\ref{fig:findings:rating2}; \control=6.70$\pm$0.46, \baseline=6.00$\pm$1.10, $z$=3.00, $p$=0.096). 

One explanation for these similar levels of satisfaction, despite the differences in creative control, maybe the unexpected (and irrelevant) yet joyful, creative discoveries participants encountered in \baseline{} condition. 
We observed from the think-aloud data that, while transforming inspirations into musical elements with \aria{} often led to off-topic suggestions, participants still \textbf{encountered moments of serendipity}. 
\aria{}'s generations were sometimes ``\textit{eye-opening}'' (P1), even if they did not fit the original plan. In these cases, participants \textbf{deviated from their initial inspiration and still arrived at satisfying results}. For instance, P8 initially set out to express `nostalgia for a childhood camp where my friends and I played card games,' but the final result was an epic heavy metal song. P8 reflected, ``\textit{It didn't feel like myself---I kind of followed the genre \aria{} gave me and adjusted my goal to suit it. I didn't feel like I had a clear idea going in. But, I'm still happy---the card games got dragons and crazy monsters in them, and these feel very important and epic from a kid's point of view.}''
This observation suggests that participants were able to derive satisfaction from both adhering to and deviating from their original creative vision, indicating that serendipitous moments of discovery can be a key factor in creative satisfaction.

% \subsubsection{\textbf{Answer to RQ3: How does \system{} impact the songwriting experience and perceived quality of the compositions?}}
\subsubsection{\textbf{Summary of findings}}
\label{section:findings:rq3:summary}
\rev{Our study reveals that \system{} positively impacts the songwriting experience by enhancing participants' agency, creativity, and perceived efficiency than using \aria{} alone. 
However, some participants experienced a sense of disconnection when switching between the tools, which could be improved by incorporating each other's control inputs.
Interestingly, despite the enhanced agency and creativity, the perceived quality of final compositions and enjoyment of the process were similar across both conditions. This similarity suggests that creative satisfaction may stem not only from control and alignment but also from unexpected discoveries throughout the process.} 
\section{Discussion}

We reflect on our findings from the design and evaluation of \system{} and discuss the lessons learned and future opportunities for music creativity support systems.

\subsection{Toward Contextual Multimodal Systems}

Overall, we found that \system{} effectively 
supported songwriters in exploring ideas while maintaining their sense of agency (\S\ref{section:findings:rq3}). 
However, our study revealed moments of friction 
when participants transitioned from the multimodal tool (\system{}) to the unimodal contextual tool (\aria{}). 
Participants who engaged with \system{} often expected \aria{} to 
generate suggestions aligned with the sources of inspiration they inputted in \system{}, only to find that \aria{} could not reflect this prior context (\S\ref{section:findings:rq3:1}). 
Moreover, while \system{} was designed to support the early stages of songwriting, four out of 10 participants used it mid-process---after they had already developed a musical structure in the editor---even though \system{} was not designed to account for existing user inputs. 
The global key conditioning of our chord generation system occasionally resulted in chord outputs that happened to be relevant to the surrounding context, 
but there were also instances where less contextually relevant outputs led to frustration (\S\ref{section:findings:rq2:control}). 

\rev{These frictions reflect a misalignment between user expectations and tool capabilities---participants formed assumptions about one tool's capability based on their experiences with the other, leading to frustration when their expectations were unmet~\citep{1908oliversatisfaction}, and consequently, a disruption in their creative flow~\citep{10.1145/3563657.3595977, csi1997creativity, csi1997flow}. 
For instance, 
interacting with \system{} during the early stages of songwriting 
likely led participants to expect that these modalities 
would continue to play a role throughout the entire songwriting process. 
When transitioning to \aria{}, however, 
their expectations clashed with \aria{}'s narrower focus on musical elements, which lacked the multimodality they had grown accustomed to.
Similarly, some participants who used \system{} mid-process expected it to consider both their multimodal inspirations and the evolving musical elements in the editor, which it did not do. 
}

Addressing these discrepancies requires establishing consistency across tools used in the creative process~\citep{norman2013design}. 
An ideal system would be both multimodal \textit{and} contextual, dynamically generating outputs that integrate both multimodal inspirations and the evolving musical context. 
This would ensure that users' creative expectations align with the system's capabilities. 
However, a key challenge in simultaneously achieving these dual goals lies in the absence of multimodal-contextual data (e.g., paired keyword-MIDI data) necessary for training models that can account for both users' multimodal and contextual inputs. 
Future work could involve building a dataset that captures the relationship between multimodal inspirations and musical elements. 
One potential approach is to collect interaction data~\citep{coauthor}, for example, by assigning specific songwriting tasks (e.g., `compose a song reflecting the mood of an image') and logging the suggestions (e.g., melodies or chords) users accept from a contextual model like \aria{}. 
Another approach might involve using multimodal models for the reverse task of generating relevant multimodal inspirations from musical elements to create paired data. 
These approaches would offer valuable multimodal-contextual data pairs to inform the development of more integrated systems.

\subsection{The Role of Randomness in Human-AI Co-Creation} 
\rev{One of the interesting observations from our study was that participants felt equally satisfied with the final compositions and enjoyed the songwriting process similarly, regardless of whether they used only \aria{} or both \aria{} and \system{}, despite \system{}’s outputs being considered as more controllable and better aligned with their creative intentions (\S\ref{section:findings:rq3:4}). 
Notably, participants often described \aria{}'s suggestions as
out-of-sync and random, yet they also reported moments of serendipity 
where these unexpected outputs prompted them to 
deviate from their original plans and still achieve satisfactory outcomes.
This observation aligns with prior work showing that 
unrelated suggestions can be evocative and productive in creative contexts~\citep{Ward_Lawson_2009, 10.1145/1142405.1142428, 10.1145/3613904.3642040}, 
with the surprising combinations of ideas 
provoking exploratory processes that lead to the production of creative ideas~\citep{kunda1990combining}.
Participants may have viewed \aria{}'s random outputs not as bugs 
but as an inherent feature of the human-AI interaction, 
and this engagement with randomness likely stimulated cognitive reorganization and creative leaps that enriched their overall experience~\citep{elephant}.

Another hypothesis is that participants' engagement styles would have played a crucial role in how they interacted with the systems and perceived their creative processes. 
As suggested by prior work in the creative writing domain~\citep{elephant}, creative processes involve broadly different styles of engagement with suggestions, including reactive and proactive engagements. 
Reactive writers let AI suggestions actively shape their work, 
using the AI outputs to determine their creative direction.
Proactive writers, by contrast, have a clear idea of what they want to create and tightly control their process. 
Similarly, participants with reactive engagement style in our study may have found \aria{}'s randomness more inspiring than \system{}'s controllable outputs, as it enabled exploration without requiring explicit decisions about the creative direction.
However, it is important to note that our study was conducted in a controlled environment, where participants were given a songwriting prompt and instructed to use the assigned tool(s) at least once.
This setup might have influenced how they engaged with the systems, potentially differing from how they would approach the tools in more natural settings.
Future work could explore these dynamics in open-ended, real-world scenarios with a larger and more diverse group of participants. 
Such studies could uncover richer usage patterns and provide deeper insights into 
how randomness, controllability, and user engagement styles contribute to 
shaping creative satisfaction and the effectiveness of AI creativity support tools.}

\subsection{Reflection on the Chord Transcriber's Low Usage}
\rev{While the majority of participants in the formative study highlighted existing music and transcribing elements like chords or melodies as a key ideation process (\S\ref{section:formative:findings}), 
the user study revealed a contrasting preference: 
participants generally favored the Chord Generator over the Chord Transcriber (\S\ref{section:findings:rq1:3}). 
Although practical constraints---such as the absence of instruments and limited time to choose appropriate songs during the study---may have influenced this preference, the difference in how the two tools supported creative exploration may provide additional insights.

We hypothesize that the Chord Generator, by accepting non-musical inputs such as images and texts, offered the creative freedom necessary to compose songs based on songwriting prompts (e.g., ``Write an 8-bar chorus about your favorite summer holiday memory.''), which was the focus of our user study task.
Working to understand the abstract and open-ended relationship between non-musical 
inputs and musical outputs likely fostered imaginative and interpretative exploration.
Theories of creative thought also support this hypothesis, which suggests that abstract thinking facilitates creative cognition by stimulating new associations and broadening the scope of possibilities~\citep{finke1996creative, ward2013creativity}. 
In contrast, the Chord Transcriber, which outputs literal transcriptions from musical inputs, provided a more concrete and constrained interaction. This specificity may limit participants’ opportunities for abstract thinking and exploratory ideation. 
As a result, the Chord Generator’s capacity to support interpretative and open-ended exploration likely made it better suited for our songwriting task, which emphasized creative ideation.
Future work could explore contexts in which more literal, concrete tools like the Chord Transcriber are particularly effective.

Additionally, participants may have preferred the Chord Generator 
because it provided multiple suggestions, 
aligning with the design principles for creativity support tools 
that emphasize the importance of exploring various alternatives~\citep{Resnick2018}. 
In comparison, the Chord Transcriber's singular, literal outputs 
may have constrained creative exploration.
Although we designed the Chord Transcriber based on observations from the formative study---where users transcribed chords from audio and expanded upon them---participants in the user study appeared to expect the tool to support further exploration of the chords, as indicated by P6 in \S\ref{section:findings:rq1:3}. 
Future work could consider expanding the Chord Transcriber's capabilities to encourage exploration within the creative space, such as offering multiple variations of transcribed chord progressions, like alterations or substitutions, alongside the literal transcriptions.}

\subsection{Real-time Multimodal Creativity Support}
Like many existing AI music systems~\citep{cococo, expressive, socialglue}, \system{} was designed for a static environment where users sit down at a computer to create music, and our experiments reflected this setting. However, music creation is inherently \textit{dynamic}, often occurring in spontaneous moments, such as during improvisation on instruments~\citep{6c2435d2-3876-39bb-b699-ca6d37615a18, walton2017creating}. 
We observed similar behaviors in both our formative and user studies. The majority of formative study participants reported they rely on improvising on their instruments to manually conjure ideas (\S\ref{section:formative:findings:transform}), 
and a user study participant noted that they would have improvised on the piano to develop ideas if they had access to it during the study session (\S\ref{section:findings:rq1:3}). 

To this end, a promising direction for the future development of human-AI music co-creation systems would involve building AI systems and interaction methods that facilitate real-time creative exploration.
One intuitive approach is real-time accompaniment systems~\citep{benetatos2020bachduet, wu2024adaptive, songdriver}, which includes harmonizing with~\citep{wu2024adaptive} or generating basslines~\citep{benetatos2020bachduet} upon user melody inputs. 
Exploring how users can effectively interact with these systems to create a meaningful composition would be a valuable area of investigation.

Moreover, multimodal systems could further enhance this process by providing real-time feedback across different modalities, supporting the user's creative exploration of musical ideas. Similar to writing support tools that offer multimodal feedback to guide story development~\citep{elephant, fairytailor}, a real-time multimodal feedback system for music could offer lyrical ideas, visual cues, or rhythmic patterns that respond dynamically to the user's improvisations, enriching the creative process.

\subsection{Expanding Multimodal Support Across Inputs and Outputs}
In this study, we focused on generating chord progressions based on multimodal inspirations. While chord progressions are a critical building block of music, the music itself comprises multiple layers: lyrics, melody, chords, bassline, drums, structure, and instrumentation, to name a few. 
A promising research direction would involve expanding multimodal support to \emph{output} other types of layers, each of which might constitute a reusable musical element. 
This could offer users greater flexibility in how they approach music composition and help them develop different layers of their songs simultaneously or in sequence. 

Similarly, musicians might benefit from AI assistance supporting diverse forms of \emph{input} \rev{beyond the text, image, and audio modalities considered in this work.} 
For instance, users could initiate their creative process with rhythm (as sensory data) or motion information (as video data). 
Exploring how different modalities can inspire musical ideas could open up new possibilities for future research.
The diversity of input/output directions pursued by participants in the AI Song Contest~\citep{aisongcontest, haisp} provides some evidence that greater flexibility on both ends would support the creative goals of more users.

\subsection{Limitations}

Several limitations of the study warrant consideration. First, \system{} always displayed the Chord Generator first, potentially influencing participant preferences. Additionally, the controlled study environment, where participants were required to use each tool at least once, may not reflect how they would naturally interact with the tools in a more open-ended setting. The novelty effect of \system{} also cannot be decoupled, as most participants were familiar with \aria{} but not with \system{}. Future studies could benefit from in-the-wild, long-term investigations to capture more natural and organic use of \system{}. 
\section{Conclusion} 
We present \system{}, a songwriting assistant that transforms multimodal (image, text, audio) inspirations into reusable musical elements (chord progressions) that can be seamlessly incorporated into songwriters' creative process. \system{} offers two key functionalities: Chord Generator and Chord Transcriber. The Chord Generator allows users to extract music keywords from multimodal inspirations and generate chord progressions aligned with these keywords. The Chord Transcriber enables users to convert audio sources into chords. To ensure the generated chord progressions are diverse, musically coherent, and relevant to keywords, we developed a novel rejection-sampling-based approach that leverages the general capabilities of LLMs in combination with a unimodal chord generation model trained on real music data.
Our technical evaluation and user study demonstrate that \system{} effectively supports the transformation of multimodal inspirations into musical elements, enhancing the songwriting process by improving perceived agency, creativity, and efficiency. With advancements in music AI, we believe \system{} opens new avenues for human-AI co-creation in music.

\begin{acks}
We thank our study participants for their valuable feedback on the design of \system{}. 
We thank
% CHI 2025 ACs and reviewers, % manuscript
Sherry Tongshuang Wu, 
Nikolas Martelaro, 
Sihyun Yu, 
DaEun Choi, 
and 
Ryuhaerang Choi for their insights, discussions, and support.
We also thank 
the Hooktheory team 
for facilitating connections with Hooktheory users and providing insightful feedback and resources for the system implementation.
% Use of AI
% https://medium.com/sigchi/acm-publications-policy-guidance-for-sigchi-venues-87332173aad1
We disclose the use of generative AI tools in the process of writing this manuscript. 
These tools were used solely for editing the authors' own text, and the authors ensured the final content was free from plagiarism, misrepresentation, fabrication, and falsification. 
% Funding
This work was supported by the National Research Foundation of Korea (NRF) grant funded by the Korea government (MSIT) (RS-2024-00337007). 
This work was in part supported by Hooktheory.
\end{acks}

\bibliographystyle{ACM-Reference-Format}
\bibliography{_references}

\newpage
\appendix 

\section{Prompts}
\label{appendix:prompts}
We list the GPT-4o prompts used in \system{} and its evaluation.

\subsection{Keywords Extraction}
\label{appendix:prompts:keywords}
\begin{framed}
    \footnotesize
    \texttt{
    \\
    \textbf{System Prompt}\\
    You are a creative assistant that generates song keywords. You will receive one or more of the following: an image URL, a text note, and user-written keywords. Your task is to create keywords that capture the essence of the input, using style-words, genres, and song-types from the provided keyword list. You may also create new relevant keywords if necessary. Respond only with the keywords, avoiding any additional commentary or formatting.\\ \\
    Instructions: \\
    1. Analyze the inputs: For an image, identify visual elements that suggest musical themes and emotions. For text, identify the main themes, emotions, or ideas. For user-written keywords, expand upon them with related music styles, genres, and types. \\
    2. Generate keywords: Use the provided keyword list, but feel free to create new, relevant keywords if they better capture the input. \\
    3. Format the output: Provide the keywords in a comma-separated list without any additional text or formatting. \\ \\
    Keyword List: \\
    Style: dance, festive, groovy, mid-tempo, syncopated, tipsy, atmospheric, cold, dark, doom, dramatic, sinister, adjunct, art, capriccio, mellifluous, nü, progressive, unusual, anthemic, emotional, happy, jubilant, melancholy, sad, aggressive, banger, power, stadium, stomp, broadway, cabaret, lounge, operatic, storytelling, torch-lounge, theatrical, troubadour, vegas, ethereal, majestic, mysterious, ambient, cinematic, heat, minimal, slow, sparse, german schlager, glam, glitter, bedroom, chillwave, intimate, sadcore, carnival, distorted, glitchy, haunted, hollow, musicbox, random, arabian, bangra, calypso, chalga, egyptian, hindustani, hōgaku, jewish music, klezmer, matsuri, middle east, polka, russian navy song, suomipop, tribal \\
    Genre: appalachian, bluegrass, country, folk, freak folk, western, afro-cuban, dance pop, disco, dubstep, disco funk, edm, electro, high-nrg, house, trance, downtempo, synthwave, trap, cyberpunk, drum'n'bass, electronic, hypnogogical, phonk, synthpop, techno, bebop, gospel, frutiger aero, jazz, latin jazz, RnB, soul, bossa nova, forró, mambo, salsa, tango, afrobeat, dancehall, dub, reggae, reggaeton, black metal, deathcore, death metal, heavy metal, heavy metal trap, metalcore, nu metal, power metal, pop, pop rock, kpop, jpop, classic rock, blues rock, emo, glam rock, hardcore punk, indie, industrial rock, punk, rock, skate rock, skatecore, funk, hiphop, rap \\
    Type: elevator, jingle, muzak, adan, adjan, call to prayer, gregorian chant, i want song, hero theme, strut, march, military, villain theme, lullaby, nursery rhyme, sing-along, toddler, adagio, andante, allegro, acoustic guitar, bass, doublebass, electricbass, electric guitar, fingerstyle guitar, percussion, chaotic, noise, stuttering, glissando trombone, legato cello, orchestral, spiccato violins, staccato viola, symphonic, 1960s, barbershop, big band, classic, doo wop, girl group, salooncore, swing, americana, christmas carol, a cappella, arabian ornamental, dispassionate, melismatic, monotone, narration, resonant, spoken word, sprechgesang, sultry, scream, torchy, vocaloid
    \\ \\
    \textbf{Examples} \\
    \textbf{User: } Image: [A base64-encoded image of a city skyline at sunset] \\
    \textbf{System: } ambient, lo-fi house moods, mellow, smooth jazz, chillwave, urban pop, dreamy, city pop \\
    \textbf{User: } Image: [A base64-encoded image of a long-distance couple reuniting at the airport] | Text Note: Met after three years at the airport. Felt like a movie scene. I scanned the crowd, and there you were, smiling. I knew I was home. \\
    \textbf{System: }cinematic, indie, nostalgic, intimate, troubadour, americana, orchestral, piano \\
    \textbf{User: }Text Note: “Hope” is the thing with feathers - That perches in the soul - And sings the tune without the words - And never stops at all. | User Keywords: hopeful \\
    \textbf{System: } emotional, ethereal, acoustic guitar, folk, atmospheric, storytelling ballad, soft-spoken harmonies
    \\
    }
\end{framed}

\subsection{Chord Progression Generation: \system{}}
\label{appendix:prompts:chordsamuse}
\begin{framed}
    \footnotesize
    \texttt{
    \\
    \textbf{System Prompt}\\
    You are a musical assistant generating chord progressions based on user-provided keywords, key, mode, and bar. The keywords describe the genre, style, and song type. The key specifies a root note that is in [C, G, D, A, E, B, F\#, Db, Ab, Eb, Bb, F]. The mode specifies a scale that is in [Maj, Min, Dor, Phr, Lyd, Mix, Loc, Hmin, Phdm]. The bar specifies the number of chords to generate for each progression. Your task is to create 30 diverse chord progressions conforming to the keywords, key, and mode. Each progression should consist of the same number of chords as the bar input, with each chord separated by a space ' ' and each progression on a new line.\\ \\ 
    Instructions: \\
    1. Analyze Chord Functions: Determine the functions of chords in the given key and mode. Tonic (I, vi) provides resolution and stability. Subdominant (IV, ii) creates movement away from the tonic. Dominant (V, vii°) creates tension that needs to resolve to the tonic.\\
    2. Analyze the Keywords: Determine the chord components and progression patterns based on the keywords. For example, for jazz-related keywords, consider using seventh chords, altered chords, and common jazz progressions like ii-V-I. For keywords like 'sadness' or 'emotional,' use minor chords, diminished chords, and progressions that create tension. \\
    3. Generate 30 Chord Progressions: Create 30 distinct chord progressions that fit the specified key and mode and match the keywords. Each progression should be unique and align with the bar parameter (i.e., if bars = 4, each progression should have 4 chords). Ensure diversity by varying the chord components (root, quality, extensions, alterations, etc.), progression patterns (diatonic/chromatic), and cadences. \\ \\
    Each chord text can have the following components, in order: \\
    1. Root Note: A-G, with optional accidentals (\#, b, x). \\
    2. Chord Quality: maj, min, aug, dim. \\
    3. Extensions: Specific chord extensions such as 6/9, 7, 9, 11, 13. \\
    4. Suspended Chords: Suspended chords such as sus2, sus4, sus\#2, sus\#4. \\
    5. Added Notes: Added notes such as add2, add4, add6, add9, add11, add13. \\
    6. Altered Notes: Alterations such as b5, \#5, b9, \#9, \#11, b13. \\
    7. Slash Chords: Alternate bass notes such as /E, /G\#, /Bb, /Dx.\\ \\
    Ensure the chord progressions are musically coherent, stylistically appropriate, and diverse. Include extensions, suspensions, adds, altered notes, slash chords as needed to achieve maximum diversity. Use both diatonic and chromatic chords to enhance the progressions. Respond only with the chord progressions, avoiding any additional commentary or formatting. \\ \\
    Examples: \\
    User keywords: dreamy, jazz, soft | Key: B | Mode: Maj | Bars: 4 \\
    Example Progressions: 
    C\#m7 F\#7 Bmaj9 d\#dim/C\textbackslash nEmaj7 A\#m7b5 D\#m7 G\#7\\ 
    User keywords: singer-songwriter, acoustic, emotional | Key: F\# | Mode: Maj | Bars: 3 \\ 
    Example progressions: 
    F\# B/F\# C\#/G\#\textbackslash nC\# D\#msus2 D\#m/ \\ 
    User keywords: orchestral, adventurous, epic | Key: D | Mode: Min | Bars: 4 \\
    Example progressions: 
    dm gm/Bb gm dm\textbackslash nBb F C C\#dim \\ \\
    Generate 30 progressions for each user input, following the above guidelines and ensuring diversity and musical coherence. Keep the chord format the same as the examples provided (e.g., G, Amaj7, Cm are valid formats, but Gmaj, Cmin are invalid formats).
    \\
    }
\end{framed}

\subsection{Chord Progression Generation: Baseline}
\label{appendix:prompts:chordsbaseline}
\begin{framed}
    \footnotesize
    \texttt{
    \\
    \textbf{System Prompt}\\
    You are a musical assistant generating chord progressions based on user-provided keywords, key, mode, and bar. The keywords describe the genre, style, and song type. The key specifies a root note that is in [C, G, D, A, E, B, F\#, Db, Ab, Eb, Bb, F]. The mode specifies a scale that is in [Maj, Min, Dor, Phr, Lyd, Mix, Loc, Hmin, Phdm]. The bar specifies the number of chords to generate for each progression. Your task is to create a chord progression conforming to the keywords, key, and mode. The progression should consist of the same number of chords as the bar input, with each chord separated by a space ' ' and the progression on a new line.\\ \\
    Instructions:\\
    1. Analyze Chord Functions: Determine the functions of chords in the given key and mode. Tonic (I, vi) provides resolution and stability. Subdominant (IV, ii) creates movement away from the tonic. Dominant (V, vii°) creates tension that needs to resolve to the tonic.\\
    2. Analyze the Keywords: Determine the chord components and progression patterns based on the keywords. For example, for jazz-related keywords, consider using seventh chords, altered chords, and common jazz progressions like ii-V-I. For keywords like 'sadness' or 'emotional,' use minor chords, diminished chords, and progressions that create tension.\\
    3. Generate a Chord Progression: Create a chord progression that fits the specified key and mode and matches the keywords. The progression should align with the bar parameter (i.e., if bars = 4, the progression should have 4 chords).\\ \\
    Each chord text can have the following components, in order: \\
    1. Root Note: A-G, with optional accidentals (\#, b, x). \\
    2. Chord Quality: maj, min, aug, dim. \\
    3. Extensions: Specific chord extensions such as 6/9, 7, 9, 11, 13. \\
    4. Suspended Chords: Suspended chords such as sus2, sus4, sus\#2, sus\#4. \\
    5. Added Notes: Added notes such as add2, add4, add6, add9, add11, add13. \\
    6. Altered Notes: Alterations such as b5, \#5, b9, \#9, \#11, b13. \\
    7. Slash Chords: Alternate bass notes such as /E, /G\#, /Bb, /Dx.\\ \\
    Ensure the chord progression is musically coherent and stylistically appropriate. Include extensions, suspensions, adds, altered notes, and slash chords as needed to achieve a rich and satisfying progression. Use both diatonic and chromatic chords to enhance the progression. Respond only with the chord progression, avoiding any additional commentary or formatting. \\ \\
    Examples: \\
    User keywords: dreamy, jazz, soft | Key: B | Mode: Maj | Bars: 4 \\
    Example Progression: 
    C\#m7 F\#7 Bmaj9 d\#dim/C\\ 
    User keywords: singer-songwriter, acoustic, emotional | Key: F\# | Mode: Maj | Bars: 3 \\ 
    Example progression: 
    F\# B/F\# C\#/G\# \\ 
    User keywords: orchestral, adventurous, epic | Key: D | Mode: Min | Bars: 4 \\
    Example progression: 
    dm gm/Bb gm dm \\ \\
    Generate a progression for each user input, following the above guidelines and ensuring musical coherence. Keep the chord format the same as the examples provided (e.g., G, Amaj7, Cm are valid formats, but Gmaj, Cmin are invalid formats).
    \\
    }
\end{framed}

\section{Chord Progression Generation Details}
\label{appendix:techdetails}

\subsection{Rejection Sampling Implementation Details}

$P(\mathbf{x})$ consists of two stacked LSTM layers with 512 embedding and hidden dimensions, a learning rate of 1e-5, and a dropout rate of 0.2. Similarly, $Q(\mathbf{x})$ uses two stacked LSTM layers but with 256 embedding and hidden dimensions, the same learning rate of 1e-5, and a dropout rate of 0.2. 
The distribution is smoothed with a temperature value of 1.7 during rejection sampling.

\subsection{Keywords Used for Listening Study}
\label{appendix:techdetails:keywords}
We list 10 sets of keywords used for the ``Keyword Relevance'' section of the listening study.

\begin{itemize}
    \item urban, hip hop, trap
    \item powerful, rock, heavy metal
    \item emotional ballad, sad
    \item bossa nova, latin jazz, samba
    \item acoustic, folk, country
    \item soul, r\&b, groovy
    \item smooth, jazz, swing
    \item lo-fi, dreamy, ambient
    \item orchestral, epic, cinematic
\end{itemize}

In addition to this, we included the attention check question with keywords ``beautiful, calm, piano'' where users select between consonance and dissonance chord progressions.

\begin{table}[t!]
    \centering\small
    \caption{\rev{CLAP scores calculated using the synthetic audio used to evaluate keyword relevance in the listening study (Section~\ref{section:techeval:rejsampling}). \system{} achieves higher scores ($\uparrow$), indicating better relevance compared to the baselines.}}
    \rev{
    \begin{tabular}{l c}
        \toprule
         Method & CLAP~$\uparrow$ \\
         \midrule
         LSTM Prior & 0.2626 \\
         GPT-4o & 0.2661 \\
         \rowcolor{aliceblue} \system{} & 0.2685 \\
         \bottomrule
        % \toprule
        % Q(X): unigram 0.37, bigram 0.53
    \end{tabular}
    }
    \label{tab:eval:clap}
\end{table}

% \system{} generates keyword-conditioned chord progressions that are more coherent, i.e., closer to real music data distributions, as indicated by Jensen-Shannon Divergence~(JSD)~\cite{jsd} between each chord generation method and music data~\cite{sheetsage}. $\downarrow$ indicates that lower values are better. We compute JSD for unigram/bigram chord distributions. LSTM Prior refers to chords generated by an LSTM trained on music data ($P(\mathbf{x})$). GPT-4o refers to chords generated by GPT-4o ($Q(\mathbf{x}|\mathbf{c})$) with keywords marginalized. \system{} represents rejection-sampled GPT-4o using LSTM Prior.

\subsection{Additional Evaluation on Keyword Relevance}
\label{appendix:techdetails:clap}

\rev{We report the automatic evaluation results for keyword relevance using the CLAP score~\citep{laionclap2023, htsatke2022} to complement the qualitative findings from our listening study (Section~\ref{section:techeval:rejsampling:human}). 
The CLAP score measures how well audio aligns with a text prompt by computing the pairwise cosine similarity between text and audio embeddings extracted by CLAP.
CLAP is a dual-encoder model, with one encoder processing text inputs and the other processing audio inputs. These embedding spaces are jointly trained using a contrastive learning objective~\citep{oord2019rep}. 
For evaluation, we feed the synthetic audio (used in the listening study for keyword relevance) into the CLAP audio encoder and set the CLAP text encoder input to ``\textit{An audio track of [keywords] music},'' building on prior work~\citep{musiccontrolnet, laionclap2023}.

Table~\ref{tab:eval:clap} presents the results: LSTM prior, GPT4o, and \system{} achieved CLAP scores of 26.3\%, 26.6\%, and 26.9\%, respectively.
Note that scores in the range of 20\%-30\% are typical for cosine similarity metrics in multimodal contrastive models (e.g., CLIP~\citep{clip})~\citep{musiccontrolnet, yu2024efficient, singer2022make, podell2023sdxl}.
In contrast, in our listening study, listeners rated those systems as having better pairwise keyword relevance 41.4\%, 50.7\%, and 58.0\% of times. 
For each pair in the listening study, we measured the correlation between differences in CLAP scores and listener judgments. This resulted in a percent agreement of 48.67\% and a Pearson correlation of 0.05 (p=0.25). 
We suspect that the CLAP audio encoder cannot meaningfully embed our synthetic audio, resulting in a weak correlation overall. 
Although CLAP scores and listener judgments follow a similar trend (favoring \system{}), the listening study provides a more reliable evaluation in this setting.}

\section{User Study}
\label{appendix:study}

\subsection{Survey Questions}
\label{appendix:study:surveyquestions}
For the post-task surveys in the user study, participants were asked to rate their agreement with the following statements on a seven-point Likert scale (1=Strongly Disagree, 7=Strongly Agree). [system(s)] was either ``Aria alone'' (\baseline) or ``\system{} and Aria together'' (\control).

\begin{itemize}
    \item \textbf{Inspiration Support}: ``Using [system(s)], I could easily translate abstract ideas (e.g., imagery, emotions) into concrete musical components.’’
    \item \textbf{Task Alignment}: ``Using [system(s)], I felt the tool(s) guided me toward a composition that felt connected to the task prompt.’’
    \item \textbf{Think Through}: ``Using [system(s)], I could think through what kinds of outputs I would want to complete the task goal, and how to complete the task.’’
    \item \textbf{Output Quality}: ``Using [system(s)], the outputs I received from the tool(s) were of sufficient quality to be useful in my composition.’’
    \item \textbf{Diversity}: ``Using [system(s)], I was able to receive a wide range of suggestions that I wouldn't have composed on my own.’’
    \item \textbf{Collaborative}: ``Using [system(s)], I felt I was collaborating with the tool(s) to come up with the composition.’’
    \item \textbf{Controllable}: ``Using [system(s)], I felt I had control over the songwriting process. I could steer the tool(s) towards the task goal.’’
    
    \item \textbf{Ownership}: ``I feel that the composition I created using [system(s)] is mine.’’
    \item \textbf{Musical Coherence}: ``I feel that the composition I created using [system(s)] is musically coherent.’’
    \item \textbf{Potential for Development}: ``I feel that the composition I created using [system(s)] could serve as a strong foundation for a full song that I could expand on.’’
    \item \textbf{Enjoyment}: ``Using [system(s)], I was very engaged in the songwriting activity - I enjoyed and would do it again.’’
    \item \textbf{Exploration}: ``Using [system(s)], it was easy for me to explore many different options, ideas, designs, or outcomes.’’
    \item \textbf{Expressiveness}: ``Using [system(s)], I was able to be very expressive and creative while doing the activity.’’
    \item \textbf{Immersion}: ``Using [system(s)], I was able to concentrate on the activity, and I forgot about the tool that I was using.’’
    \item \textbf{Results Worth Effort}: ``Using [system(s)], what I was able to produce was worth the effort required to produce it.’’
\end{itemize}

The questions for think-through, collaborative, and controllable are adapted from the previous work that measured user perceptions on AI systems~\cite{aichains}. The questions for ownership and musical coherence are adapted from the related work on human-AI music co-creation~\cite{expressive}. The questions for enjoyment, exploration, expressiveness, immersion, and results worth effort are Creativity Support Index~\cite{csi} measures.

\subsection{Interview Questions}
\label{appendix:study:interviewquestions}
We list the questions used for the semi-structured interview after the two songwriting sessions with baseline and \system{}.

\begin{itemize}
    \item Can you walk me through your songwriting process when using both Aria and \system{} together?
    \item Comparing the baseline and \system{}, what were the main differences you noticed in the songwriting process?
    \item How did using \system{} with Aria influence your final composition compared to using Aria alone? Did it lead to any unexpected or particularly satisfying results?
    \item Were there any creative challenges you faced during the process that the tools helped (or could have helped) you overcome? How did they assist (or not) in solving specific problems?
    \item Which features in \system{} did you find most helpful or not? Can you describe specific scenarios where they were particularly helpful or unhelpful?
    \item How do you envision using tools like Aria and \system{} in your future songwriting projects? What role do you see them playing in your creative process?
    \item What specific improvements or features would you suggest for either Aria, \system{}, or both to better support your songwriting process? Are there any gaps or frustrations you encountered that you want addressed?
\end{itemize}

\subsection{Thematic Analysis Codebook}
\label{appendix:study:codebook}
\rev{We include the codebook used for the thematic analysis of user experience on \system{} and \aria{}. Table~\ref{tab:appendix:codebook1} presents the codes related to user experiences in the \control{} condition where users used both \system{} and \aria{}. Table~\ref{tab:appendix:codebook2} presents the codes related to user experiences in the \baseline{} condition where users used \aria{} alone.}

\begingroup
\setlength{\tabcolsep}{8pt} % Default value: 6pt 
\begin{table*}[t]

\caption{\rev{Codebook summarizing the dimensions, descriptions, and example quotes from participants regarding the benefits, challenges, and feature usability of \system{} integrated with \aria{} in the \control{} condition.}}
\label{tab:appendix:codebook1}

\centering
\renewcommand{\arraystretch}{1.25} 
\scalebox{0.9}{ % adjust the scale factor as needed
\rev{ % Change font color to blue
\begin{tabular}{p{3.5cm}p{4.5cm}p{7cm}}
\toprule
 \textbf{Dimensions and Codes} & \textbf{Code Description} & \textbf{Example Quote} \\ 
    \midrule
    \multicolumn{2}{l}{\textit{What are the benefits of using \system{} with \aria{}?}} \\
    \dashedline{1-3}

    Controllable outcomes & Users can steer the tool outputs to align with their creative intentions.  & ``\textit{\system{} was definitely more helpful when starting from scratch. The ideas it generated fell in line with I would have mentally constructed.}'' \\ 
    
    Diverse outcomes & The tools generate novel and diverse outputs, providing inspirations or surprises to users. & ``\textit{\system{} surprised me with its use of chromaticism, pushing me to think beyond diatonic patterns. It's been fun and engaging to work with ideas I wouldn't have come up with on my own.}'' \\

    Foster imagination & The tools facilitate imaginative thinking and creative expression. & 
    ``\textit{\system{} did allow for a more generative ideation. I felt more creatively connected to what I was actually doing musically from a very imaginative perspective.}'' \\

    Streamline workflow & Users find creation process efficient and easy with the tools. & ``\textit{I really liked how it jump-started the process. I usually spend a lot of time and effort getting the right chords, which is tedious---being able to convey an idea quickly was really nice.}'' \\

    Agency over workflow & User has a sense of autonomy in their interactions with the tools. & ``\textit{It felt like unlocking a communication channel where I can tell what I wanted, and compared to AI writing the whole thing for me, with \system{}, I had control in the writing process.}'' \\
    
    \midrule
    \multicolumn{2}{l}{\textit{What are the challenges of using \system{} with \aria{}?}} \\
    \dashedline{1-3}

    \makecell[tl]{Disconnection\\ between tools} & Switching between \system{} and \aria{} disrupts the creative workflow. & ``\textit{The only disconnection I felt was in how \aria{} doesn’t necessarily integrate with \system{} in the same way that \system{} integrates with \aria{}. So it was almost like doing a 180-degree rotation when switching from \system{} to \aria{} because \aria{} doesn’t achieve quite the same results, given that it doesn’t consider your inputs.}'' \\

    \midrule
    \multicolumn{2}{l}{\textit{Which features in \system{} did users particularly helpful or unhelpful?}} \\
    \dashedline{1-3}

    Music keywords are helpful & Users find music keywords in Chord Generator helpful. & ``\textit{Turning an image directly into chords feels like a big abstract leap. Generating keywords from the image helps me connect with it---asking, `What keywords come to mind for this image?' makes the process feel more familiar and predictable.}'' \\

    Image input is helpful & Users find image input in Chord Generator helpful. & ``\textit{I especially liked the picture input feature because, as the saying goes, a picture is worth a thousand words. Sometimes, it's easier to convey what you're aiming for with an image than with words.}''\\

    \makecell[tl]{Chord Transcriber \\ is unhelpful} & Users find Chord Transcriber unnecessary or unhelpful. & ``\textit{I didn't use the transcribe feature which I think mainly is because of the time constraint for me---I couldn't come up with a song that kind of fit the task that I could borrow from off the top of my head.}'' \\
    
    \bottomrule
\end{tabular}
} % end text color
}
\vspace{1em}
\end{table*}
\renewcommand{\arraystretch}{1} 
\endgroup
\begingroup
\setlength{\tabcolsep}{8pt} % Default value: 6pt 
\begin{table*}[t]

\caption{\rev{Codebook summarizing the dimensions, descriptions, and example quotes from participants regarding the benefits and challenges of using \aria{} alone in the \baseline{} condition.}}

\centering
\renewcommand{\arraystretch}{1.25} 
\scalebox{0.9}{ % adjust the scale factor as needed
\rev{ % font color
\begin{tabular}{p{3.5cm}p{4.5cm}p{7cm}}
\toprule
 \textbf{Dimensions and Codes} & \textbf{Code Description} & \textbf{Example Quote} \\ 

    \midrule
    \multicolumn{2}{l}{\textit{What are benefits of using \aria{} alone?}} \\
    \dashedline{1-3}

    Intuitive workflow & Users find the standalone \aria{} experience straightforward. & ``\textit{I would say when \system{} was there, I felt more enriched and musical, but using \aria{} alone was more intuitive because I knew what to expect from it.}'' \\

    Unexpected discoveries & Users receive unexpected, serendipitous results that inspire new directions from \aria{}. & ``\textit{I feel like \aria{} generates something randomly so it could or could not work. But in this case, the fact that it worked and it actually sounded good was pretty nice experience.}'' \\

    \midrule
    \multicolumn{2}{l}{\textit{What are the challenges of using \aria{} alone?}} \\
    \dashedline{1-3}

    Uncontrollable outcomes & Users find \aria{}'s output random or irrelevant, failing to align with the user's creative intentions. & ``\textit{\aria{} is a bit random -- its suggestions don't sync up exactly with what I am writing.}''\\

    \makecell[tl]{Lack of agency \\ over workflow} & Users feel they lack autonomy in their interactions with \aria{}. & ``\textit{With just \aria{}, I wasn’t able as much to create a mood or feel for my song. I didn’t feel like I had much input—\aria{} just sort of made up its own mind, and I had to just go along with it. It’s like having a bad co-writer.}''\\

    \bottomrule
\end{tabular}
} % font color 
}
\label{tab:appendix:codebook2}
\vspace{-.5em}
\end{table*}
\renewcommand{\arraystretch}{1} 
\endgroup

\end{document}